\documentclass[aps,onecolumn,showpacs]{revtex4}  % for review and submission
\usepackage{graphicx}  % needed for figures
\usepackage{dcolumn}   % needed for some tables
\usepackage{bm}        % for math
\usepackage{amssymb}   % for math
\usepackage{amsmath}
\usepackage{dsfont}

\newcommand{\tr}{{\textrm{Tr}\,}}
\newcommand{\be}{\begin{eqnarray}}
\newcommand{\ee}{\end{eqnarray}}
\newcommand{\pp}[1]{\phantom{#1}}

\begin{document}

\title{Theory versus experiment for vacuum Rabi oscillations in lossy cavities}
\author{ Marcin Wilczewski and Marek Czachor}
\affiliation{
Katedra Fizyki Teoretycznej i Informatyki Kwantowej\\
Politechnika Gda\'nska, 80-952 Gda\'nsk, Poland\\
and\\
Centrum Leo Apostel (CLEA)\\
Vrije Universiteit Brussel, 1050 Brussels, Belgium
}

\begin{abstract}
The 1996 Brune {\it et al.\/} experiment on vacuum Rabi oscillation is analyzed by means of alternative models of atom-reservoir interaction. Agreement with experimental Rabi oscillation data can be obtained if one defines jump operators in the dressed-state basis, and takes into account thermal fluctuations between dressed states belonging to the same manifold. Such low-frequency transitions could be ignored in a closed cavity, but the cavity employed in the experiment was open, which justifies our assumption. The cavity quality factor corresponding to the data is approximately $Q=3.31\cdot 10^{10}$, whereas $Q$ reported in the experiment was $Q=7\cdot 10^7$. The rate of decoherence arising from opening of the cavity can be of the same order as an analogous correction coming from finite time resolution $\Delta t$ (formally equivalent to collisional decoherence). Peres-Horodecki separability criterion shows that the rate at which the atom-field state approaches a separable state is controlled by fluctuations between dressed states from the same manifold, and not by the rate of transitions towards the ground state. In consequence, improving the $Q$ factor we do not improve the coherence properties of the cavity.
\end{abstract}

\pacs{42.50.Lc, 42.50.Dv, 32.80.Ee, 32.80.Qk}
\maketitle

%\section{\label{sec:level1}First-level heading}
% sections are not used for PRL papers
\section{Introduction}

In 1996 the group from Laboratoire Kastler--Brossel in Paris reported to date most precise measurements of vacuum Rabi oscillations in cavity QED \cite{Brune}. The suggestive plots from \cite{Brune} have achieved status of classic textbook illustrations of agreement between theory and experiment in quantum optics \cite{Schleich,Gerry}. The experimental setup is close to ideal, and as such is a promising candidate for implementation of quantum logic gates. This brings, however, all the issues one will encounter in a realistic quantum computer, with qubit-reservoir interactions and various kinds of decoherence in the first place.

When one tries to reconstruct the data from theoretical models one encounters surprising and intriguing difficulties. As far as we know, none of the authors who undertook this task managed to find a convincing explanation of all the discrepancies. The main goal of our paper is to discuss theoretical consequences of the fact that the cavity used in the experiment was open. We will see that then
at least one new important source of nondissipative decoherence, not taken into account so far, arises.

The first attempt of explaining the data on the basis of master equations was the PhD thesis of Y. T. Chough, written already a year after the experiment but, unfortunately, not widely known \cite{ChoughPhD,Chough}. Chough identified two main effects that did not find support in his quantum-trajectory simulations: Too large amplitude damping (probability of atomic ground state was too quickly decaying towards 1/2) and too small energy damping (the probability did not quickly enough approach 1). Chough showed that the data are much better approximated by a semiclassical theory. The problem with both kinds of damping was discussed from a general model-independent perspective by Bonifacio {\it et al.\/} \cite{Bonifacio}. Here the conclusion was that the amplitude decay could not be caused by energy losses. The authors localized one possible non-dissipative source of the damping, namely experimental time uncertainty $\Delta t$ following from atomic velocity distribution. Indeed, neglecting energy losses but taking into account uncertainties in the measurement of $t$, one observes that Rabi oscillation of the ground-state probability gets damped towards 1/2.

Yet another intriguing possibility was noticed by ourselves in \cite{Czachor}. Our original intention was to check if vacuum Rabi oscillation experiments are capable of distinguishing between alternative approaches to field quantization. The first observation we made was again the problem with too strong amplitude damping. Since in this context this was precisely the effect we hoped to see (fields quantized in reducible non-Fock representations of canonical commutation relations(CCR) \cite{Cachor1,Czachor2,Czachor3,CzachorNaudts,CzachorWrzask} should lead in this situation to vacuum collapses and revivals) we made some effort to discriminate between different sources of such behavior. It turned out that the result was inconclusive: The effects discussed in \cite{Bonifacio} and \cite{Czachor} can mask one another, unless the uncertainties in $\Delta t$ are sufficiently small and one can monitor the oscillation long enough to see the revival --- if it really occurs. However, the lack of energy decay noticed by Chough was clearly seen also in the alternative formalism, so in this respect the situation was not better than in the earlier approaches. It became clear that a more reliable theory of dissipation is needed before one can say anything concrete about the problem as fundamental as the choice of representation of CCR employed in field quantizaton.

Relatively recently Scala {\it et al.\/} \cite{Scala} questioned the very basics of master equations employed in the quantum optics literature, and proposed an apparently cosmetic modification of the standard formalism. We will describe this approach in detail in the following sections, so let us sketch here the main result, which turns out to go back to the original ideas of Davies on how to describe system-reservoir interactions  in Markovian master equations \cite{Davies,Davies2}. The observation is very simple but essential for general consistency of the dynamics: The ``jump operators" should describe jumps between eigenstates of the system Hamiltonian (i.e. dressed states), and not eigenstates of the free-field subsystem (free-photon number states) as it happens in practically all approaches employed in quantum optics \cite{Alicki}.

We will see that an  open-cavity generalization of the model from \cite{Scala} we propose allows for reconstruction of experimental data even in the idealized case of infinite time resolution, $\Delta t=0$. The generalized construction involves jump operators defined in the dressed-state basis but, as opposed to this from \cite{Scala}, takes into account also the jumps (up and down) between dressed states belonging to the same manifold. Such jumps cannot be excluded in open cavities since at $T=0.8$K average numbers of appropriate long-wave photons are very large. The cavity of Brune {\it et al.\/} was open so the effect has to be included. Of course, $\Delta t$ was nonzero and collisions with background gases were present. As the latter two effects are mathematically equivalent, we decided to supplement our generalized model by the general scheme from \cite{Bonifacio}. In consequence, we seem to have included the three main ingredients of decoherence: Collisions and uncertainty of $t$, losses of energy, and the external long-wave modes. However, even now we obtain agreement with the data only under the assumption that the cavity $Q$ factor was much higher than assumed in \cite{Brune}, so some  of the problems noticed in \cite{ChoughPhD,Chough} do not disappear. A tentative interpretation of this discrepancy is that the additional Davies operator we have introduced may arise from system reservoir interactions that are sensitive to photon number \cite{Alicki} (cf. the Appendix). One more observation we make is that theoretical treatments given in \cite{ChoughPhD,Chough,Bonifacio,Czachor} did not properly distinguish between the notions of ``true" and ``effective" times, somewhat imprecisely discussed in \cite{Brune}, and this led to incorrect estimates of decay parameters of the cavity in question.

The two types of transitions we consider (within a single manifold or between different manifolds of dressed states) turn out to have different physical consequences. For example, these are the ``ignored" fluctuations within a single manifold that control separability properties of the atom--photon system, and this is the rate of this transition that plays a role of the damping parameter known from Bloch equation treatments of Rabi oscillations.

Following the terminology from \cite{Scala} we will speak of standard quantum optical ``phenomenological" master equations, while the one of Scala {\it et al.\/} will be termed ``microscopic". We will confront the data with predictions of both models at $T=0$K and $T=0.8$K, and then formulate the open-cavity model at $T=0.8$K. In comparison with the data we assume exact resonance. Not only will this assumption simplify formulas, but it leads to closed-form solutions for a cavity with a Gaussian mode profile. What is interesting, the latter problem can be computed exactly in our generalized model, but in the phenomenological formalism we had to resort to numerical solutions. So the approach we regard as more physical, is simultaneously mathematically simpler.

\section{Phenomenological model at zero temperature and exact resonance}
\label{sec:phenomenological-model}

One begins with the master equation
\begin{eqnarray}
\dot{\rho}
&=&%1
-i [\Omega,\rho]
+
\gamma
\big(
a \rho a^\dagger
-
\frac 1 2
[a^\dagger a, \rho]_+
\big),
\label{eq:me-stdT0}
\end{eqnarray}
where $[A,B]_+=AB+BA$.
The von Neumann part of the dynamics is governed by the Jaynes-Cummings Hamiltonian $H=\hbar\Omega$, where
\begin{eqnarray}
\Omega
&=&%1
\frac{\omega_0}{2}
\left( |e\rangle \langle e| -  |g\rangle \langle g|  \right)
+
\omega_0
a^\dagger a
+
{\texttt{g}}
\left(
a |e\rangle \langle g|
+
a^\dagger
|g\rangle \langle e|
\right).
\label{eq-hjc}
\end{eqnarray}
Creation and annihilation operators $a$, $a^\dagger$ occurring in $\Omega$ play also a dual role of Davies jump operators \cite{Davies,Davies2} in the dissipative part of the master equation. The coupling parameter ${\texttt{g}}$ is, for simplicity, assumed to be real.
$\hbar\omega_0$ is the energy separation between the atomic upper $|e\rangle$ and lower $|g\rangle$ states. The decay parameter $\gamma$ is responsible for energy losses.

Let us note already at this stage that this particular model of dissipation assumes that the losses occur only for states with non-zero photon numbers: The atom-cavity system loses energy through transitions $|g,1\rangle\to|g,0\rangle$, but $|e,0\rangle\to|g,0\rangle$ is not allowed. The dissipator in equation (\ref{eq:me-stdT0}) could be supplemented  by an additional term describing atomic spontaneous emission at its natural rate unmodified by the cavity. Within the cavity QED regime, however, this term can be neglected. In this sense, the presence of the atom inside of the cavity is practically ignored in the phenomenological model.

In the following discussion we will concentrate on dynamics that starts from the pure state $|e,0\rangle \langle e,0|$. The assumption implies a convenient constraint on the Hilbert space in question. Indeed, at $T=0$K the system can make transitions only downwards on the energy ladder, so that the only states we should take into account are: $|1\rangle= |e,0\rangle$, $|2\rangle = |g,1\rangle$ and $|3\rangle=|g,0\rangle$. In this basis Eq.~(\ref{eq:me-stdT0}) is equivalent to the following three sets of coupled first-order differential equations
\begin{subequations}
\begin{eqnarray}
\dot{\rho}_{11}
&=&%1
i {\tt{g}}  \rho_{12}
-
i {\tt{g}}  \rho_{21},
\label{eq:61}
\\
\dot{\rho}_{22}
&=&%2
-
\gamma \rho_{22}
-
i {\tt{g}} \rho_{12}
+
i {\tt{g}} \rho_{21},
\label{eq:62}
\\
\dot{\rho}_{12}
&=&%3
-
\frac{ \gamma }{ 2 }
\rho_{12}
+
i {\tt{g}} \rho_{11}
-
i {\tt{g}} \rho_{22},
\label{eq:63}
\\
\dot{\rho}_{21}
&=&%4
-
\frac{ \gamma }{ 2 }
\rho_{21}
-
i {\tt{g}} \rho_{11}
+
i {\tt{g}} \rho_{22},
\label{eq:64}
\\
\dot{\rho}_{33}
&=&%5
\gamma
\rho_{22},
\label{eq:65}
\end{eqnarray}
\end{subequations}
\begin{subequations}
\begin{eqnarray}
\dot{\rho}_{13}
&=&%1
-i \omega_0 \rho_{13} - i  {\tt{g}} \rho_{23},
\label{eq:66}
\\
\dot{\rho}_{23}
&=&%2
- i \omega_0 \rho_{23} - \frac{ \gamma }{ 2 } \rho_{23} - i {\tt{g}} \rho_{13},
\label{eq:67}
\end{eqnarray}
\end{subequations}
and
\begin{subequations}
\begin{eqnarray}
\dot{\rho}_{31}
&=&%1
i \omega_0 \rho_{31} + i  {\tt{g}} \rho_{32}
\label{eq:68}
\\
\dot{\rho}_{32}
&=&%2
i \omega_0 \rho_{32} - \frac{ \gamma }{ 2 } \rho_{32} + i {\tt{g}} \rho_{31}.
\label{eq:69}
\end{eqnarray}
\end{subequations}
The loss rate $\gamma$ does not affect the population $\rho_{11}$, while it contributes to the decay of $\rho_{22}$ and the increase of $\rho_{33}$. In other words, the population of the state $|e,0\rangle$ is not directly affected by the dissipation, in contrast to the case of $|g,1\rangle$ (dissipation decreases it) and $|g,0\rangle$ (dissipation increases it).  The coherences $\rho_{12}$ and $\rho_{21}$ are also damped but with the rate $\gamma/2$.

If we assume that the atom-field system is initially prepared in the state $|e,0\rangle$, the initial condition is $\rho_{11}(0)=1$, with all the other elements vanishing. With these conditions the last two sets of equations have solutions identically equal to zero,
\begin{eqnarray}
\rho_{13}(t)
=
\rho_{31}(t)
=
\rho_{23}(t)
=
\rho_{32}(t)
=
0.
\end{eqnarray}
The remaining five equations imply
\begin{subequations}
\begin{eqnarray}
\rho_{11}(t)
&=&%1
-
\frac{  8 {\texttt{g}}^2  }{\gamma ^2-16 {\texttt{g}}^2}
e^{-\frac{ \gamma }{ 2 }  t}
+
\frac{ \gamma   }{2 \sqrt{\gamma ^2-16   {\texttt{g}}^2}}
e^{-\frac{ \gamma }{ 2 }  t}
\left(e^{\frac{1}{2} t \sqrt{\gamma ^2-16 {\texttt{g}}^2}}-e^{-\frac{1}{2} t \sqrt{\gamma ^2-16 {\texttt{g}}^2}}\right)
\nonumber
\\
&&%2
+
\frac{ \gamma^2 - 8 {\texttt{g}}^2 }{2 \left(\gamma ^2-16 {\texttt{g}}^2\right)}
e^{- \frac{ \gamma }{2}  t}
\left(e^{\frac{1}{2} t \sqrt{\gamma ^2-16
   {\texttt{g}}^2}}+e^{-\frac{1}{2} t \sqrt{\gamma ^2-16 {\texttt{g}}^2}}\right),
\label{eq:rho11}\\
\rho_{22}(t)
&=&%1
\frac{ 4 {\texttt{g}}^2 }{\gamma ^2-16 {\texttt{g}}^2}
e^{-\frac{\gamma }{2} t}
\left(e^{\frac{1}{2} t \sqrt{\gamma ^2-16 {\texttt{g}}^2}}+e^{-\frac{1}{2} t \sqrt{\gamma ^2-16 {\texttt{g}}^2}}\right)
-
\frac{ 8 {\texttt{g}}^2  }{\gamma ^2-16 {\texttt{g}}^2}
e^{-\frac{ \gamma }{2} t }
\label{eq:rho22},\\
\rho_{33}(t)
&=&%1
1
+
\frac{ 16 g^2 }{\gamma ^2-16 g^2}
e^{-\frac{ \gamma  }{2} t }
-
\frac{ \gamma ^3  }{2 \left(\gamma ^2-16 g^2\right)^{3/2}}
e^{-\frac{ \gamma }{2} t}
\left(e^{\frac{1}{2} t \sqrt{\gamma ^2-16 g^2}}-e^{-\frac{1}{2} t \sqrt{\gamma ^2-16 g^2}}\right)
\nonumber
\\
&&%2
-
\frac{\gamma ^2 }{2 \left(\gamma ^2-16 g^2\right)}
 e^{-\frac{ \gamma }{2} t}
\left(e^{\frac{1}{2} t \sqrt{\gamma ^2-16 g^2}}+e^{-\frac{1}{2} t \sqrt{\gamma ^2-16 g^2}}\right)
\nonumber
\\
&&%3
+
\frac{  8 g^2 \gamma   }{\left(\gamma ^2-16 g^2\right)^{3/2}}
e^{-\frac{ \gamma }{ 2 } t}
\left(e^{\frac{1}{2} t \sqrt{\gamma ^2-16 g^2}}-e^{-\frac{1}{2}t \sqrt{\gamma ^2-16 g^2}}\right)
\label{eq:rho33},\\
\rho_{12}(t)
&=&%1
-
2 i
\frac{  {\texttt{g}}  \gamma    }{\gamma ^2-16{\texttt{g}}^2}
e^{-\frac{ \gamma }{ 2 } t}
+
i
\frac{  {\texttt{g}} }{\sqrt{\gamma ^2-16   {\texttt{g}}^2}}
\left( e^{\frac{1}{2} t \sqrt{\gamma ^2-16{\texttt{g}}^2}} - e^{-\frac{1}{2} t \sqrt{\gamma ^2-16{\texttt{g}}^2}}   \right)
e^{-\frac{ \gamma }{ 2 }  t}
\nonumber
\\
&&%2
+
i
\frac{  \gamma {\texttt{g}}}{\gamma^2-16{\texttt{g}}^2}
\left(e^{\frac{1}{2} t \sqrt{\gamma ^2-16{\texttt{g}}^2}}  + e^{-\frac{1}{2} t \sqrt{\gamma ^2-16{\texttt{g}}^2}}  \right)
e^{-\frac{ \gamma }{ 2 }  t},
\label{eq:rho12}\\
\rho_{21}(t)
&=&%1
2 i
\frac{  {\texttt{g}}  \gamma    }{\gamma ^2-16{\texttt{g}}^2}
e^{-\frac{ \gamma }{ 2 } t}
-
i
\frac{  {\texttt{g}} }{\sqrt{\gamma ^2-16   {\texttt{g}}^2}}
\left( e^{\frac{1}{2} t \sqrt{\gamma ^2-16{\texttt{g}}^2}} - e^{-\frac{1}{2} t \sqrt{\gamma ^2-16{\texttt{g}}^2}}   \right)
e^{-\frac{ \gamma }{ 2 }  t}
\nonumber
\\
&&%2
-
i
\frac{  \gamma {\texttt{g}}}{\gamma^2-16{\texttt{g}}^2}
\left(e^{\frac{1}{2} t \sqrt{\gamma ^2-16{\texttt{g}}^2}}  + e^{-\frac{1}{2} t \sqrt{\gamma ^2-16{\texttt{g}}^2}}  \right)
e^{-\frac{ \gamma }{ 2 }  t}.
\label{eq:rho21}
\end{eqnarray}
\end{subequations}
We will be particularly interested in the probabilities $p^{\text{ph}}_{e,0}(t)=\rho_{11}(t)$ , $p^{\text{ph}}_{g,1}(t)=\rho_{22}(t)$, and $p^{\text{ph}}_{g,0}(t)=\rho_{33}(t)$. Noting that the quantity $\gamma^2 - 16 \texttt{g}^2$ appearing in all the above equations takes negative values within the strong coupling regime ($\texttt{g}$ is much larger than atomic and cavity relaxation times) we get
\begin{subequations}
\begin{eqnarray}
p^{\text{ph}}_{e,0}(t)
&=&%1
\frac{  8 {\texttt{g}}^2  }{16 {\texttt{g}}^2 - \gamma ^2}
e^{-\frac{ \gamma }{ 2 }  t}
+
\frac{ \gamma   }{\sqrt{ 16   {\texttt{g}}^2  -  \gamma ^2 }}
e^{-\frac{ \gamma }{ 2 }  t}
\sin   \frac t 2 \sqrt{16 \texttt{g}^2 -\gamma ^2}
+
\frac{ 8 {\texttt{g}}^2 - \gamma^2   }{16 {\texttt{g}}^2 - \gamma ^2}
e^{- \frac{ \gamma }{2}  t}
\cos \frac t 2   \sqrt{16 \texttt{g}^2 - \gamma ^2},\\
p^{\text{ph}}_{g,1}(t)
&=&%1
\frac{ 8 {\texttt{g}}^2  }{16 {\texttt{g}}^2 - \gamma ^2}
e^{-\frac{ \gamma }{2} t }
-
\frac{ 8 {\texttt{g}}^2 }{16 {\texttt{g}}^2 - \gamma ^2}
e^{-\frac{\gamma }{2} t}
\cos \frac t 2 \sqrt{16 \texttt{g}^2 - \gamma ^2},\\
p^{\text{ph}}_{g,0}(t)
&=&%1
1
-
\frac{ 16 \texttt{g}^2 }{16 \texttt{g}^2 - \gamma ^2}
e^{-\frac{ \gamma  }{2} t }
+
\frac{ \gamma ^3  }{ \left( 16 \texttt{g}^2 -  \gamma ^2  \right)^{3/2}}
e^{-\frac{ \gamma }{2} t}
\sin \frac t 2  \sqrt{16 \texttt{g}^2 - \gamma^2}
\nonumber
\\
&&%2
+
\frac{\gamma ^2 }{16 \texttt{g}^2   -  \gamma ^2}
 e^{-\frac{ \gamma }{2} t}
\cos \frac t 2  \sqrt{16 \texttt{g}^2 -  \gamma ^2}
-
\frac{  16  \texttt{g}^2 \gamma   }{\left( 16 \texttt{g} ^2 - \gamma ^2 \right)^{3/2}}
e^{-\frac{ \gamma }{ 2 } t}
\sin \frac t 2 \sqrt{16 \texttt{g}^2 - \gamma ^2}.
\end{eqnarray}
\end{subequations}
Asymptotically, for $t\to\infty$, we find
$
\lim_{t\to\infty}
p^{\text{ph}}_{g,0}(t)
=
1
$
and
$
\lim_{t\to\infty}
p^{\text{ph}}_{g,1}(t)
=0$.
We conclude that the photon must finally escape from the cavity. The solid curves at Fig.~1 show the evolution of $p^{\text{ph}}_{g,0}(t)$ and $p^{\text{ph}}_{g,1}(t)$
for the atom-field coupling characteristic of the Brune {\it et al.\/} experiment \cite{Brune}. The damping parameter $\gamma$ is at this stage of our analysis chosen arbitrarily, just to show the essential qualitative properties of the dynamics.

\section{Phenomenological model at $T>0$ --- First approximation}

Let us now proceed to a more complex and at the same time more realistic case of $T>0$. The phenomenological master equation acquires an additional term related to upwards transitions on the energy ladder,
\begin{eqnarray}
\frac{d \rho}{dt}
&=&%1
-i
[\Omega,\rho]
+
\gamma_\downarrow
\Big(
a \rho a^\dagger
-
\frac 1 2
[a^\dagger a,\rho]_+
\Big)
+
\gamma_\uparrow
\Big(
a^\dagger \rho a
-
\frac 1 2
[a a^\dagger,\rho]_+
\Big).
\label{eq:me-stdT}
\end{eqnarray}
Here, $\gamma_\downarrow$ and $\gamma_\uparrow$ denote rates of transitions downwards and upwards, respectively. Given temperature $T$, these two coefficients are related by the formula (a consequence of the KMS condition \cite{KMS})
\begin{eqnarray}
\gamma_\uparrow
&=&%1
e^{- \frac{\hbar \omega_0 }{ k T}}
\gamma_\downarrow.
\end{eqnarray}
Inserting $T$ and $\omega_0$ as reported in the experiment \cite{Brune} ($T=0.8$K and $\omega_0=2 \pi\, 51.099$GHz) we obtain
\begin{eqnarray}
\gamma_\uparrow
& = &%1
\epsilon
\gamma_\downarrow,
\qquad
\text{ with } \epsilon\approx 0.0466327,
\end{eqnarray}
which, in fact, means that Eq. (\ref{eq:me-stdT}) contains a single parameter $\gamma_\downarrow$. The most important formal difference with respect to $T=0$ is that now the number of states that take part in the evolution becomes infinite, no matter what initial condition one takes. This is a consequence of two facts: The form of the thermal density matrix at $t=0$, and the possibility of upwards transitions implied by $\gamma_\uparrow\neq 0$.
However, for $T=0.8$K the contribution to the initial condition from all non-vacuum states is not greater than $5\%$ of probability, and the upward transition to higher dressed states is of the same order. So the technical assumption we make here is that in the first approximation it is legitimate to truncate the Hilbert space to its three-dimensional subspace spanned by $|1\rangle=|e,0\rangle$, $|2\rangle=|g,1\rangle$ and $|3\rangle=|g,0\rangle$, but including the upwards transitions within this subspace. The next approximation would be to include the next dressed-state subspace, but within this subspace include only the downwards transitions --- this correction will be discussed in a separate paper.

Once we make the simplifying assumption, we have to modify the form of the Davies operators to guarantee that higher dimensional subspaces of the Hilbert space will not couple to the subspace in question. Let $\Pi$ be the projector on the subspace of interest,
\begin{eqnarray}
\Pi
&=&%1
|e,0\rangle \langle e,0|
+
|g,1\rangle \langle g,1|
+
|g,0\rangle \langle g,0|,
\end{eqnarray}
so that the truncated Davies operators are obtained by the substitution
\begin{eqnarray}
a
&\to&%1
\Pi
a
\Pi
=
|g,0\rangle \langle g,1|
,
\\
a^\dagger
&\to&%2
\Pi
a^\dagger
\Pi
=
|g,1\rangle \langle g,0|.
\end{eqnarray}
The modified master equation (\ref{eq:me-stdT}) now reads
\begin{eqnarray}
\dot{\rho}
&=&%1
-i[\Omega,\rho]
\nonumber
\\
&&%2
+
\gamma_\downarrow
\Big(
|g,0\rangle \langle g,1| \rho |g,1\rangle \langle g,0|
-
\frac 1 2
[|g,1\rangle \langle g,1|, \rho]_+
\Big)
\nonumber
\\
&&%2
+
\gamma_\uparrow
\Big(
|g,1\rangle \langle g,0| \rho |g,0\rangle \langle g,1|
-
\frac 1 2
[|g,0\rangle \langle g,0|, \rho]_+
\Big).
\end{eqnarray}
As we can see, the dissipative term again does not affect the state $|e,0\rangle$ of the system. Matrix elements of $\dot{\rho}$ between basis states split into three groups of equations,
\begin{subequations}
\begin{eqnarray}
\dot{\rho}_{11}
&=&%1
i {\tt{g}}  \rho_{12}
-
i {\tt{g}}  \rho_{21},
\label{eq:3030}
\\
\dot{\rho}_{22}
&=&%2
-
\gamma_\downarrow \rho_{22}
-
i {\tt{g}} \rho_{12}
+
i {\tt{g}} \rho_{21}
+
\gamma_\uparrow \rho_{33},
\label{eq:3031}
\\
\dot{\rho}_{12}
&=&%3
-
\frac{ \gamma_\downarrow }{ 2 }
\rho_{12}
+
i {\tt{g}} \rho_{11}
-
i {\tt{g}} \rho_{22},
\label{eq:3032}
\\
\dot{\rho}_{21}
&=&%4
-
\frac{ \gamma_\downarrow }{ 2 }
\rho_{21}
-
i {\tt{g}} \rho_{11}
+
i {\tt{g}} \rho_{22},
\label{eq:3033}
\\
\dot{\rho}_{33}
&=&%5
\gamma_\downarrow
\rho_{22}
-
\gamma_\uparrow \rho_{33},
\label{eq:3034}
\end{eqnarray}
\label{eq:3035}
\end{subequations}
\begin{subequations}
\begin{eqnarray}
\dot{\rho}_{13}
&=&%1
-i \omega_0  \rho_{13}
-
i {\tt{g}} \rho_{23}
-
\frac{ \gamma_\uparrow }{ 2 }
\rho_{13},
\label{eq:100}
\\
\dot{\rho}_{23}
&=&%2
-i \omega_0 \rho_{23} -i {\tt{g}} \rho_{13} - \frac{ \gamma_\downarrow + \gamma_\uparrow }{ 2 } \rho_{23}
\label{eq:101},
\end{eqnarray}
\label{eq:102}
\end{subequations}
and
\begin{subequations}
\begin{eqnarray}
\dot{\rho}_{31}
&=&%1
i \omega_0  \rho_{31}
+
i {\tt{g}} \rho_{32}
-
\frac{ \gamma_\uparrow }{ 2 }
\rho_{31},
\label{eq:103}
\\
\dot{\rho}_{32}
&=&%2
i \omega_0 \rho_{32}
+
i {\tt{g}} \rho_{31} - \frac{ \gamma_\downarrow + \gamma_\uparrow }{ 2 } \rho_{32},
\label{eq:104}
\end{eqnarray}
\label{eq:105}
\end{subequations}
It is obvious that for $\gamma_\uparrow=0$ the equations reduce to (\ref{eq:61})-(\ref{eq:69}).
The situation is analogous to $T=0$. Assuming that the atom-field system starts from $\rho(0)=|e,0\rangle \langle e,0|$ we find that (\ref{eq:102}) and (\ref{eq:105}) again have trivial solutions $\rho_{13}(t)=\rho_{31}(t)=\rho_{23}(t)=\rho_{32}(t)=0$, so that it is sufficient to focus attention on (\ref{eq:3035}). Unfortunately, although the number of equations is the same as for $T=0$, we are not able to determine analytically the eigenvectors of the matrix of coefficients appearing at the right hand side of (\ref{eq:3035}). Numerical analysis shows, however, that predictions are basically the same as their counterparts for $T=0$K, as shown in Fig.~1. Just to have a feel of the role of $\epsilon$ for the evolution we plot in Fig.~2 the appropriate probabilities for several different $\epsilon$s. The range of temperatures we consider makes this exercise purely formal, but it clearly shows the tendency for lowering the curves as $\epsilon$ grows.

\begin{center}
\begin{figure}[h]
\includegraphics{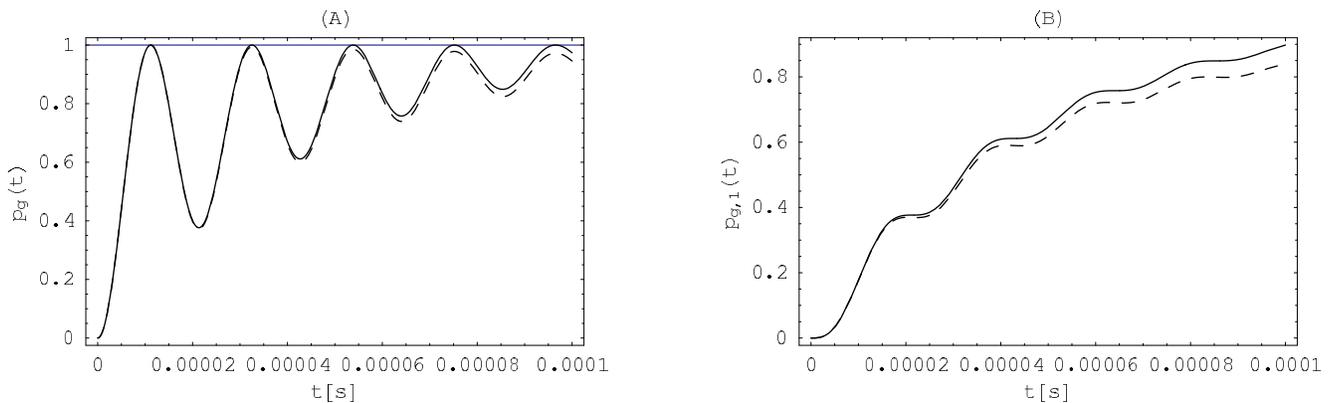}%
\caption{Predictions of the phenomenological model for $T=0$K (solid) and $T=0.8$K (dashed) for $\texttt{g}=47 \pi 10^3$Hz. The dissipation rate for $T=0$ is $\gamma=0.3 \texttt{g}$, while  $\gamma_\downarrow=0.3 \texttt{g}$ and $\gamma_\uparrow=0.0466 \gamma_\downarrow$ for $T=0.8$K.  Plot (A) shows the probability of finding the atom in its ground state, $p^{\text{ph}}_{g}(t)=p^{\text{ph}}_{g,0}(t)+p^{\text{ph}}_{g,1}(t)$, (B) shows the probability for a photon to be outside of the cavity, $p^{\text{ph}}_{g,0}(t)$.}
\label{fig:ph_modelT}
\end{figure}

\end{center}

\begin{center}
\begin{figure}[h]
\includegraphics{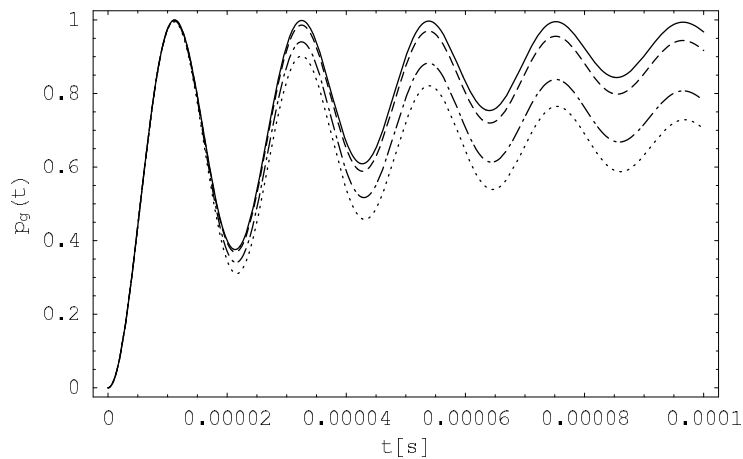}%
\caption{Predictions of the phenomenological model for $\epsilon=0.01$ (solid), $\epsilon=0.1$ (dashed), $\epsilon=0.5$ (dot-dashed) and $\epsilon=1$ (dotted) and $\texttt{g}=47 \pi 10^3$Hz, $\gamma_\downarrow=0.3 \texttt{g}$. As mentioned in the text the restriction to the three dimensional subspace is a gross abuse since the corresponding temperatures are, respectively, $T\approx 0.54$K, $T\approx 1.1$K, $T\approx 3.5$K and $T=\infty$.}
\label{fig:ph_model_epsilon}
\end{figure}
\end{center}

\section{Description in dressed-state basis --- Scala et al. approach}\label{sec:Scala-model}

The main drawback of the phenomenological model lies in the choice of states determining the possible transitions during the energy loss. These are not the eigenstates of the {\it system\/} Hamiltonian $H$, but those  of the free-field Hamiltonian $\hbar\omega a^{\dag}a$ (i.e. $|g,1\rangle$ and $|g,0\rangle$). However, in all general theories of system--reservoir interactions one assumes that the losses are modeled by transitions between the eigenstates of the Hamiltonian of the system \cite{Davies,Davies2}. In our case $H$ is the Jaynes-Cummings Hamiltonian representing bound-states of the atom--photon (or spin--oscillator) composite system. Atoms and photons are here as inseparable as protons and electrons in atoms and --- similarly to atoms --- should be assumed to loose energy {\it via\/} transitions between appropriate energy levels. The phenomenological model is implicitly based on a {\it classical\/} intuition that the system necessarily loses energy by means of losses of light, since it is light that interacts with the cavity walls, and not the atom itself (for a detailed discussion of assumptions behind the phenomenological model cf. \cite{Scala2}).

An implication for our atom--photon system is that the Davies jump operators $a$ and $\Pi a\Pi$ should be replaced by appropriate Davies operators defined in the dressed-state basis. This is the starting point of the ``microscopic model" developed by Scala {\it et al.\/} in \cite{Scala} (further generalizations beyond Markovian and rotating-wave approximations were given in \cite{Scala2,Scala3,Scala4}).
At $T=0$K the master equation proposed in \cite{Scala} reads
\begin{eqnarray}
&&%1
\dot{\rho}
=
-i [\Omega,\rho]
+
\gamma_1
\Big(
\frac 1 2
|\Omega_0\rangle \langle \Omega_+| \rho |\Omega_+\rangle \langle \Omega_0|
-
\frac 1 4
[|\Omega_+\rangle \langle \Omega_+|,\rho]_+
\Big)
+
\gamma_2
\Big(
\frac 1 2
|\Omega_0\rangle \langle \Omega_-| \rho |\Omega_-\rangle \langle \Omega_0|
-
\frac 1 4
[|\Omega_-\rangle \langle \Omega_-|,\rho]_+
\Big),
\label{me-scala}
\end{eqnarray}
where $|\Omega_\pm\rangle$ are the dressed atomic states, $|\Omega_0\rangle=|g,0\rangle$, while $\gamma_1$ and $\gamma_2$ denote rates at which the corresponding transitions $|\Omega_\pm\rangle\to |\Omega_0\rangle $ take place. Let us note that $\Omega=H/\hbar$ is the same as in the phenomenological model: We do not modify the system, but only the way it interacts with the reservoir.

When comparing this equation with (\ref{eq:me-stdT0}) or (\ref{eq:me-stdT}) one notices that it contains two jump operators, $|\Omega_0 \rangle\langle \Omega_+ |$ and $|\Omega_0 \rangle  \langle \Omega_- |$ expressed in terms of the dressed states, and both of them  describe transitions towards the system ground state $|\Omega_0\rangle=|g,0\rangle$. Within this model, therefore, both bare states $|g,1\rangle$ and $|e,0\rangle$ may decay towards $|g,0\rangle$ contributing to the irreversible loss of energy from the system. In fact, since we are now working with the dressed states which are combinations of the bare ones, even a single decay channel produces decays of the two non-ground bare states: $|e,0\rangle$ and $|g,1\rangle$. Recall, that the essence of the phenomenological model, on the other hand, is that  the only state that may directly decay is $|g,1\rangle$.

From a formal point of view, the essence of microscopic treatments of dissipation is in the way the Davies  operators are constructed. In \cite{Scala} one gets
\begin{eqnarray}
A(\Omega_{N',l}-\Omega_{N,m})
&=&|\Omega_{N,m}\rangle \langle \Omega_{N,m}| \big( a+a^\dagger \big) |\Omega_{N',l}\rangle \langle \Omega_{N',l}|
\nonumber
\\
&=&%3
\frac 1 2
\delta_{N,N'-1}
\big( \sqrt{N+1} + lm \sqrt{N} \big) |\Omega_{N,m} \rangle \langle \Omega_{N+1,l}|,
\label{scala-jump-operators}
\end{eqnarray}
where $l,m=\pm$ and $N$ refers to the total number of excitations in the atom-cavity system. One feature, to which we shall refer in the following, is that operators derived this way can describe transitions between neighboring manifolds of dressed states, but not within a single manifold. So the model excludes transitions between $|\Omega_+\rangle$ and $|\Omega_-\rangle$. This makes sense in closed cavities since the wavelength of an appropriate transition is of the order of kilometers, while the cavity sizes are much smaller (a few centimeters). However, the cavity used by Brune {\it et al.\/} was not closed --- we shall return to this issue later.

\begin{center}
\begin{figure}[t]
\includegraphics{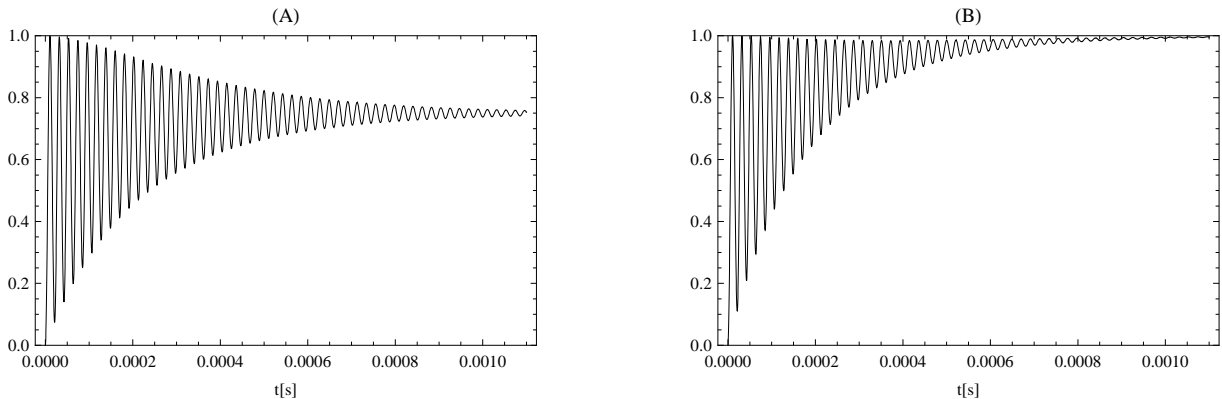}%
\caption{Predictions of the microscopic approach. Both plots show the probability of finding the atom in its ground state under different conditions: $\texttt{g}= 47 \pi 10^3$Hz; (A) $\gamma_1=0.1 \texttt{g}$, $\gamma_2=0$, and  (B)  $\gamma_1=0.1 \texttt{g}$, $\gamma_2=0.05 \texttt{g}$.}
\label{fig:scala-model}
\end{figure}
\end{center}

Assume as before that the initial state of the atom-cavity system is $|e,0\rangle$. In the basis $|1\rangle=|\Omega_+\rangle$, $|2\rangle=|\Omega_-\rangle$ and $|3\rangle=|\Omega_0\rangle$ the  master equation takes the form
\begin{subequations}
\begin{eqnarray}
\dot{\rho}_{11}
&=&%1
-
\frac{ \gamma_1 }{ 2 } \rho_{11},
\label{eq:11}
\\
\dot{\rho}_{22}
&=&%2
-\frac{ \gamma_2 }{ 2 } \rho_{22},
\label{eq:12}
\\
\dot{\rho}_{33}
&=&%3
\frac{ \gamma_1 }{2 } \rho_{11} + \frac{ \gamma_2 }{2} \rho_{22},
\label{eq:13}\\
\dot{\rho}_{12}
&=&%1
\left(  -i(\Omega_+ - \Omega_-) - \frac{ \gamma_1  + \gamma_2 }{ 4 }   \right)
\rho_{12},
\label{eq:20}
\\
\dot{\rho}_{13}
&=&%2
\left(  -i(\Omega_+ - \Omega_0) - \frac{ \gamma_1 }{ 4 }   \right)
\rho_{13},
\label{eq:21}
\\
\dot{\rho}_{23}
&=&%3
\left(  -i(\Omega_- - \Omega_0) - \frac{ \gamma_2 }{ 4 }  \right)
\rho_{23}.
\label{eq:22}
\end{eqnarray}
\end{subequations}
As we can see the equation is much simpler than the phenomenological one. The solutions are
\begin{subequations}
\begin{eqnarray}
\rho_{11}(t)
&=&%1
e^{-\frac{ \gamma_1 }{ 2 } t}
\rho_{11}(0),
\label{eq:55}
\\
\rho_{22}(t)
&=&%2
e^{-\frac{\gamma_2}{2} t}
\rho_{22}(0),
\label{eq:56}
\\
\rho_{33}(t)
&=&%3
\Big(  1 - e^{\frac{\gamma_1}{2}  t}  \Big) \rho_{11}(0)
+
\Big( 1 - e^{-\frac{\gamma_2 }{2} t} \Big) \rho_{22}(0)
+
\rho_{33}(0),\\
\label{eq:57}
\rho_{12}(t)
&=&%1
e^{\big( -i (\Omega_+ - \Omega_- ) - \frac{\gamma_1  + \gamma_2 }{4}   \big) t} \rho_{12}(0),
\label{eq:31}
\\
\rho_{13}(t)
&=&%1
e^{\big( -i (\Omega_+ - \Omega_0) - \frac{\gamma_1 }{4 }  \big)t} \rho_{13}(0),
\label{eq:32}
\\
\rho_{23}(t)
&=&%3
e^{\big(  -i (\Omega_- - \Omega_0) - \frac{ \gamma_2 }{4}  \big)t} \rho_{23}(0).
\label{eq:33}
\end{eqnarray}
\end{subequations}
In exact resonance the initial condition satisfies
\begin{eqnarray}
\rho(0)
&=&%1
|e,0\rangle \langle e,0|
\ = \
\frac 1 2 |\Omega_+\rangle \langle \Omega_+|
+
\frac 1 2 |\Omega_-\rangle \langle \Omega_-|
-
\frac 1 2
|\Omega_+\rangle \langle \Omega_-|
-
\frac 1 2
|\Omega_-\rangle \langle \Omega_+|
\label{eq:40},
\end{eqnarray}
and the solution can be written as
\begin{eqnarray}
\rho(t)
&=&%1
\frac 1 2
e^{- \frac{\gamma_1 }{ 2 } t}
|\Omega_+\rangle \langle \Omega_+|
+
\frac 1 2
e^{- \frac{\gamma_2 }{2 } t}
|\Omega_-\rangle \langle \Omega_-|
+
\Big(
1-\frac 1 2 e^{-\frac{\gamma_1 }{2 } t} - \frac 1 2 e^{-\frac{\gamma_2 }{2 }t}
\Big)
|\Omega_0\rangle \langle \Omega_0|
\nonumber
\\
&&%2
-
\frac 1 2 e^{2 i \texttt{g} t} e^{-\frac{\gamma_1 + \gamma_2 }{4} t}
|\Omega_-\rangle  \langle \Omega_+|
-
\frac 1 2 e^{-2 i \texttt{g} t} e^{-\frac{\gamma_1 + \gamma_2 }{4} t}
|\Omega_+\rangle  \langle \Omega_-|.
\label{eq:rho_scala}
\end{eqnarray}
The corresponding probabilities of the bare states are
\begin{eqnarray}
p^{\text{mic}}_{g,0}(t)
&=&%1
1
-
\frac 1 2
e^{- \frac{ \gamma_1 }{ 2 } t}
-
\frac 1 2
e^{- \frac{ \gamma_2 }{ 2 } t},\\
p^{\text{mic}}_{g,1}(t)
&=&%1
\frac 1 4
e^{- \frac{ \gamma_1 }{ 2 } t}
+
\frac 1 4
e^{- \frac{ \gamma_2 }{ 2 } t}
-
\frac 1 2
e^{- \frac{ \gamma_1 + \gamma_2 }{4} t}
\cos 2 {\textrm{g}} t.
\end{eqnarray}
The probability $p^{\text{mic}}_g(t)$ of finding the atom in the lower state
\begin{eqnarray}
p^{\text{mic}}_g(t)
&=&%1
p^{\text{mic}}_{g,0}(t)+p^{\text{mic}}_{g,1}(t)
=
1
-
\frac 1 4
e^{-\frac{\gamma_1 }{2} t}
-
\frac 1 4
e^{- \frac{ \gamma_2 }{ 2 }t }
-
\frac 1 2
e^{ - \frac{ \gamma_1 + \gamma_2 }{ 4 }  t}
\cos 2 \texttt{g} t
\end{eqnarray}
is symmetric with respect to $\gamma_1$ and $\gamma_2$, and it is the sum of these coefficients that controls the rate at which oscillations are damped. An interesting situation (population trapping) takes place when one sets $\gamma_1=0$ or $\gamma_2=0$. In this case, the probability of finding the atom in the lower state approaches asymptotically $3/4$ instead of $1$. In contrast, for non-zero $\gamma_1$ and $\gamma_2$ the limiting value is 1, as shown in Fig.~\ref{fig:scala-model}.

In the following Sections we will try to confront the above two models with the experiment reported in \cite{Brune}, and then propose yet another modification of the master equation, appropriate for an open cavity.

\section{The experiment of Brune {\it et al.\/} --- role of Gaussian profile of the cavity field}\label{sec:gaussian_in_general}

The above calculations were performed under the assumption that the atom-field coupling $\texttt{g}$ was constant in time. Still, in real experiments the coupling is effectively time dependent since the atoms are moving through the cavity. A convenient formulation of the problem is to work in the atomic rest frame, where the moving cavity field can be modeled by a time-dependent coupling parameter $\texttt{g}(t)$ \cite{Walther}.

The cavity used in this particular experiment is an open Fabry-Perot resonator consisting of two mirrors of diameter $d=50$mm each, and with mirror spacing $L=27$mm measured on the cavity axis. The mode sustained by the cavity has a Gaussian profile characterized by the waist $w=5.96$mm. The walls of the cavity (mirrors) are cooled to approximately $0.8$K which corresponds to the average number  $\overline{n}=0.05$ of photons of resonant frequency. The two-level system that interacts with the cavity field is implemented by circular Rydberg states of rubidium, with principal quantum numbers $n=50$ and $n=51$, playing the roles of  lower $|g\rangle$ and upper $|e\rangle$ levels, respectively. The transition frequency between these states is $\omega_0\approx 2\pi \thickspace 51.099$ GHz and the inverse radiative lifetime $1/\tau\approx 0.03$ kHz. The cavity relaxation time is $T_{\rm cav}=220\mu$s and $\texttt{g}=47 \pi 10^3$ Hz, hence the parameters fulfill conditions of the strong-coupling regime.
\begin{center}
\begin{figure}[t]
\includegraphics{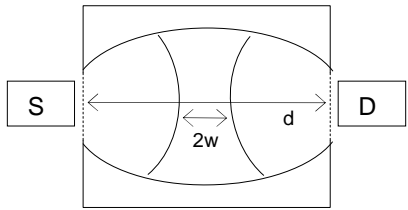}%
\caption{The experimental setup we consider. The source of atoms (S) and the detector (D) are placed, respectively, directly before and after the cavity, so that effects of dissipation outside of the cavity are neglected. The diameter of the cavity equals $d$ and duration of the evolution is determined by $t=d/v$, where $v$ is the atomic velocity.
}
\label{fig:cavity_scheme}
\end{figure}
\end{center}

The experiment is schematically shown in Fig.~4. If the time of flight between the source and the detector is $t$, the atom arrives at the center of the cavity at $t'=t/2$. The relation between the distance $d$ and the time $t$ is given by $d=vt$, where $v$ is the velocity of the atom. The run of the experiment that measures the atomic state at time $t$ corresponds to the Hamiltonian $H(t')$ whose coupling is given by
$\texttt{g}(t')=\texttt{g}e^{-v^2(t/2-t')^2/w^2}$, and the evolution runs for $0\leq t'\leq t$. Let us note that coupling during the interaction is almost all the time smaller than the maximal value $\texttt{g}=\texttt{g}(t/2)$. One expects therefore that the Rabi oscillation (at exact resonance) is not performed with the Rabi frequency $\texttt{g}$ but with some smaller effective frequency. Brune {\it et al.\/} assume that this frequency should be computed according to
\be
\texttt{g}_{\rm eff} t=\int_0^t\texttt{g}(t')dt'\approx \texttt{g}\int_{-\infty}^\infty e^{-(vt')^2/w^2}dt'=\texttt{g} \sqrt{\pi} \frac w v=\texttt{g} \sqrt{\pi} \frac w d t=\texttt{g}\,t_{\rm eff}.\label{teff}
\ee
We will later see that the recipe (\ref{teff}) works indeed for system-reservoir interactions involving the dressed-state Davies operators, but in fact is not exactly true for the phenomenological model.

\begin{center}
\begin{figure}[t]
\includegraphics{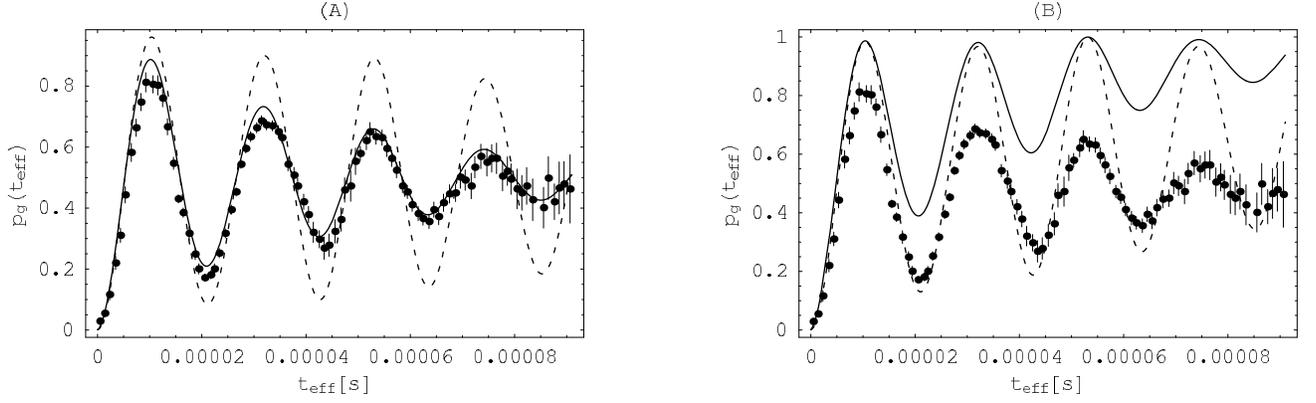}%
\caption{Probability of the atomic ground state as a function of the effective time $t_{\rm eff}$. Dots with error bars represent experimental points. (A) The dashed line represents (\ref{eq:3360}) with $\gamma^{-1}=220 \cdot 10^{-6}$s; the solid line corresponds to the corrected \mbox{formula (\ref{eq:3400})}, hence after unifying the notion of time. (B) We have simply multiplied the probabilities known from the ideal Jaynes-Cummings model by exponential factors $\exp (-\gamma t_{\rm eff})$ (dashed line) and $\exp \big(-\frac{\gamma}{\sqrt{\pi} w/d} t_{\rm eff}\big) $ (solid line). In both cases $\texttt{g}=47 \pi 10^3$Hz and $T=0.8$K.}
\label{fig:brune_experiment}
\end{figure}
\end{center}

Both the book \cite{Haroche} and the review \cite{Raimond-colloquium} make it very clear that the data reported in \cite{Brune} were plotted as functions of the effective time $t_{\rm eff}$, and not as those of the true evolution time $t$. The data obtained in the experiment are presented in Fig.~\ref{fig:brune_experiment}.
Dots with error bars represent experimental points corresponding to the probability of finding the atom in the lower state $|g\rangle$. The solid and dashed curves represent fits by means of the function
\begin{eqnarray}
p^{\text{expt}}_g(t_{\rm eff})
&=&%1
1
-
\frac 1 2
\sum_{n=0}^\infty
P(n)
\big(
1 + e^{-\gamma t_{\rm eff}} \cos(2 \texttt{g} \sqrt{n+1} \ t_{\rm eff})
\big),
\label{eq:3360}
\end{eqnarray}
where $P(n)$ is the thermal distribution, and for two different damping parameters $\gamma$.

The fitting formula (\ref{eq:3360}) has caused a lot of confusion in the literature, and there were two main reasons for that. One reason was that the first paper \cite{Brune} did not clearly state that the time axis involved $t_{\rm eff}$ and not $t$. The second reason can be understood after having solved master equations with time dependent Hamiltonians and our $\texttt{g}(t')$. We will see that inclusion of the Gaussian structure of $\texttt{g}(t')$ effectively changes $\texttt{g}t$, typical of the constant coupling, into $\texttt{g}_{\rm eff}t=\texttt{g}\,t_{\rm eff}$, but does not affect the terms $\gamma t$. It follows that if we plot the experimental data in the true time $t$, the fitting function should read
\begin{eqnarray}
p^{\text{expt}}_g(t)
&=&%1
1
-
\frac 1 2
\sum_{n=0}^\infty
P(n)
\big(
1 + e^{-\gamma t} \cos(2 \texttt{g}_{\rm eff} \sqrt{n+1} \ t)
\big).
\label{eq:3360'}
\end{eqnarray}
But if we decide to rescale the time from $t$ to $t_{\rm eff}$, as it was done in \cite{Brune}, then the fitting formula becomes
\begin{eqnarray}
\tilde p^{\text{expt}}_g(t_{\rm eff})
&=&%1
1
-
\frac 1 2
\sum_{n=0}^\infty
P(n)
\big(
1 + e^{- \frac{ \gamma }{ \sqrt{\pi} w/d }  t_{\rm eff}} \cos(2 \texttt{g} \sqrt{n+1} \ t_{\rm eff})
\big).
\label{eq:3400}
\end{eqnarray}
In other words, the effective damping parameter is $\gamma /(\sqrt{\pi} w/d )$ and not $\gamma$. Many authors, including ourselves \cite{Bonifacio,ChoughPhD,Chough,Czachor}, have noticed that the data presented in \cite{Brune} seem to correspond to a cavity whose lifetime is 40-50$\mu$s, and not $220\mu$s as claimed in \cite{Brune}. It looked like the cavity was five times worse than it was assumed. However, when we insert the cavity parameters into $\gamma /(\sqrt{\pi} w/d )$ we find that the effective lifetime of the cavity is around $46\mu$s, in agreement with the data. In Fig.~\ref{fig:brune_experiment}A the time axis corresponds to $t_{\rm eff}$. The dashed line is the fit with $\gamma t_{\rm eff}$ where $1/\gamma =220\mu$s; the solid curve is damped by  $\gamma /(\sqrt{\pi} w/d ) t_{\rm eff}$, with the same value of $\gamma$. As we can see, an explanation of {\it this\/} discrepancy may be trivial.

One should keep in mind that, in spite of this simple explanation of the ``discrepancy", the fitting formulas cannot be regarded as theoretical curves, and it is easy to understand why. The long-time asymptotics of both (\ref{eq:3360'}) and (\ref{eq:3400}) is
\be
\lim_{t_{\rm eff}\to\infty}\tilde p^{\text{expt}}_g(t_{\rm eff})=\lim_{t\to\infty}p^{\text{expt}}_g(t)=\frac 1 2,
\ee
whereas we know that the two theories we have discussed so far predict
\be
\lim_{t\to\infty}p^{\text{ph}}_g(t)=\lim_{t\to\infty}p^{\text{mic}}_g(t)> \frac 1 2,
\ee
the right-hand-side depending on damping parameters and $T$, but in all cases the limiting values being much greater than 1/2. The predictions of the two models can be mimicked by a different kind of fitting function, shown in Fig.~\ref{fig:brune_experiment}B. Physically, the two fits describe exponential damping of two different quantities: The atomic inversion in Fig.~\ref{fig:brune_experiment}A and the probability of atomic excited state in Fig.~\ref{fig:brune_experiment}B. It is the later that agrees with theories based on Markovian master equations, simultaneously disagreeing with the data. Hence the question remains if one can explain the fitting formulas of Brune {\it et al.} on a basis of master equations.

The first question we have to discuss is the precise role of the Gaussian profile of the coupling. We will begin with the microscopic model where the form of $\texttt{g}_{\rm eff}$ will be derived in an exact way. Then we turn to the phenomenological model where an exact closed formula will not be available, but the corresponding plot will be produced numerically.

\section{Models with Gaussian mode profile}

Let us begin with a time {\it independent\/} coupling constant $\texttt{g}$ occurring in the master equation $\dot \rho(t)={\cal L}(\texttt{g})\rho(t)$. The density matrix solution can be formally written as
\be
\rho(t)=e^{{\cal L}(\texttt{g})(t-t_0)}\rho(t_0).
\ee
Now let $\texttt{g}(t)$ be a discrete $n$-step approximation of the Gaussian, as shown in Fig.~6. Then
\be
\rho(t)=e^{{\cal L}(\texttt{g}_n)\Delta t}\dots e^{{\cal L}(\texttt{g}_1)\Delta t}\rho(0)\label{n step}
\ee
where $\Delta t=t/n$,  $\texttt{g}_j=\texttt{g}(t_j)$, $t_j\in [(j-1)\Delta t,j\Delta t]$,  $j\in\{1,\dots ,n\}$. In exact resonance the basis of dressed states is independent of $\texttt{g}_j$. The solution (\ref{eq:55})--(\ref{eq:33}) thus depends on coupling constants only {\it via\/} exponents of the off-diagonal elements. This allows us to compute the explicit formula valid for the Gaussian,
\begin{eqnarray}
\rho(t)
&=&%1
\left(
\begin{array}{ccc}
\frac 1 2 e^{-\frac{\gamma_1 }{2} t}
&
-\frac 1 2
e^{- \frac{\gamma_1 + \gamma_2 }{4}  t}
e^{- 2 i \texttt{g} \sqrt{\pi} \frac{w}{d}t  }
&
0\\
-\frac 1 2
e^{-  \frac{\gamma_1  + \gamma_2 }{4}  t}
e^{ 2 i \texttt{g} \sqrt{\pi} \frac{w}{d}  t  }
&
\frac  1 2
e^{-\frac{\gamma_2 }{2} t}
&
0 \\
0
&
0
&
\frac 1 2
\left( 1 - e^{-\frac{\gamma_1 }{ 2 }t} \right)
+
\frac 1 2
\left(1-e^{-\frac{\gamma_2 }{2 }t}\right)
\end{array}
\right),
\label{eq:3470}
\end{eqnarray}
by means of the limit $n\to\infty$.  The coupling parameter $\texttt{g}$ gets indeed replaced by $\texttt{g}_{\rm eff}=\texttt{g} \sqrt{\pi} w/d$, as suggested in \cite{Brune}, but the exponents involving $\gamma_1$ and $\gamma_2$ remain unchanged.
The probability of the ground state is
\begin{eqnarray}
p^{\text{mic}}_{g}(t)
&=&%3
1
-
\frac 1 4 e^{-\frac{\gamma_1 }{ 2 }t}
-
\frac 1 4 e^{-\frac{\gamma_2 }{2 }t}
-
\frac 1 2
e^{-  \frac{\gamma_1  + \gamma_2 }{4}  t}
\cos 2 \texttt{g} \sqrt{\pi} \frac{w}{d}  t.
\label{eq:3460}
\end{eqnarray}
\begin{center}
\begin{figure}[t]
\includegraphics[width=0.4\textwidth]{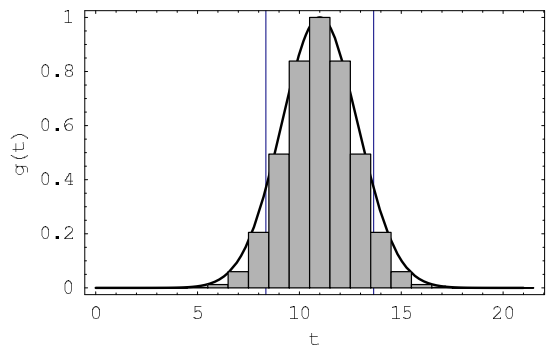}%
\caption{The interaction time is split here into $n=21$ equally-sized time intervals of length $\Delta t=t/21$. $\texttt{g}(t)$ is expressed in units of  $\texttt{g}$ and units of time are arbitrary. The distance between the two vertical lines in the central part of the plot is $2w$. The shape of the Gaussian corresponds to the mode profile inside of the cavity discussed in \cite{Brune}. In the actual numerical simulation of the phenomenological model we use $n=20001$.}
%\label{fig:cavity_scheme}
\end{figure}
\end{center}
\begin{center}
\begin{figure}[t]
\includegraphics{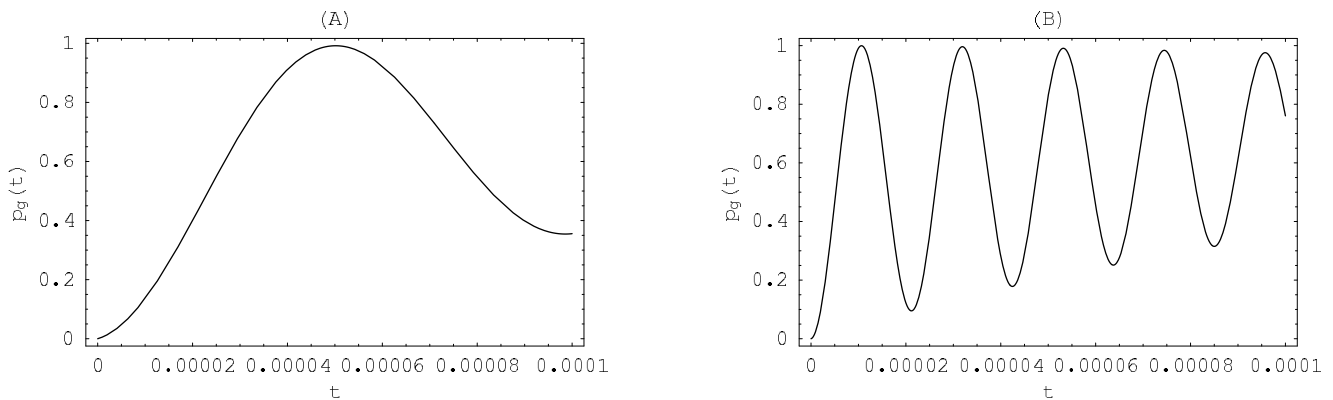}%
\caption{Predictions of the microscopic model at $T=0$, (A) with and (B) without the Gaussian structure of the coupling. In (A) the atom most of the time experiences coupling much weaker from the maximal value.  Parameters are $\texttt{g}=47 \pi 10^3$ Hz, $\gamma_1=0.12 \texttt{g}$, and $\gamma_2=0.01 \texttt{g}$. $t$ is the true time.}
\label{fig:scala_with_g_and_experiment}
\end{figure}
\end{center}
\begin{center}
\begin{figure}[t]
\includegraphics{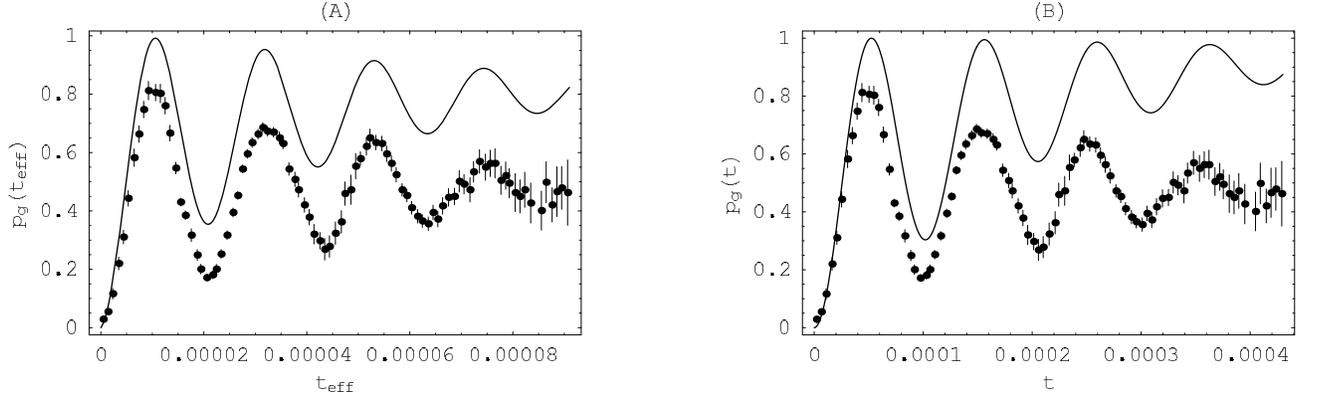}%
\caption{(A) Predictions of the microscopic model with the Gaussian coupling at $T=0$ as a function of $t_{\rm eff}$. The  parameters are $\gamma_1=0.12\texttt{g}$ and $\gamma_2=0.01\texttt{g}$. (B) Predictions of the phenomenological model with the $n$-step approximation ($n=20001$) of the Gaussian coupling at $T=0.8$K, $w=5.96$mm, $\texttt{g}=47 \pi 10^3$Hz. $\Delta t=t/20001$ for each time point (for $0\leq t\leq 430\mu$s with $1\mu$s step). $t$ is the true time. The effective Rabi frequency is somewhat smaller than $\texttt{g}_{\rm eff}$ implied by the microscopic model.  }
\label{fig:phenomenological_with_g_and_experiment}
\end{figure}
\end{center}
Fig.~\ref{fig:scala_with_g_and_experiment}A shows how the effective Rabi frequency obtained for a Gaussian coupling differs from the one found for a constant $\texttt{g}=\max_t \{\texttt{g}(t)\}$ (Fig.~\ref{fig:scala_with_g_and_experiment}B).

The case of the phenomenological model turns out to be more complicated, not only because at $T>0$ we did not manage to solve analytically the underlying eigenvalue equation, but also because even at $T=0$ the coupling constants enter the solutions (\ref{eq:rho11})--(\ref{eq:rho21}) in a very complicated way. So even at $T=0$ we would have to rely on numerical analysis. The case $T=0.8$K is shown in Fig.~\ref{fig:phenomenological_with_g_and_experiment}. Having no access to the closed-form solution of the master equation we do not know how to switch between true and effective times. So we decided to rescale the experimental data to the true time. This is why $0\leq t\leq 430\mu$s in Fig.~\ref{fig:phenomenological_with_g_and_experiment}B, as opposed to the other plots where $0\leq t_{\rm eff}\leq 91\mu$s.

Disagreement between theory and experiment is here at least as bad as in the microscopic model. The structure of solutions of these two models does not allow for fitting the data by any choice of the parameters.

\section{Open-cavity generalization of the Scala {\it et al.\/} model}

\begin{center}
\begin{figure}
\includegraphics{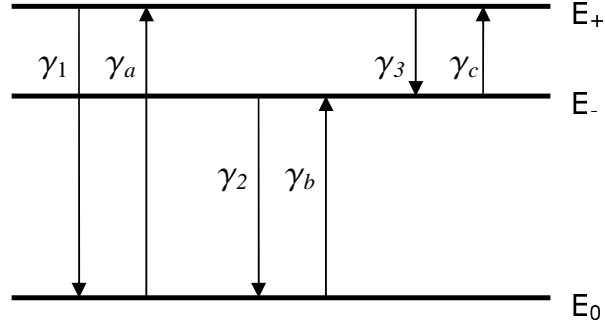}%
\caption{Energy levels and decay coefficients used in the generalized model; $E_\pm=\hbar\Omega_\pm$, $E_0=\hbar\Omega_0$.}
\label{fig:drabinka}
\end{figure}
\end{center}
At $T=0.8$K the average number of photons with energy $\Omega_+-\Omega_-=2{\texttt{g}}$ is $\bar n=354666$, but the corresponding wavelength is 6378.56 m. If the 5 cm cavity of Brune {\it et al.\/} was a closed one, there would be no reason to take $2{\texttt{g}}$ transitions into account. However, in open cavities \cite{Weinstein} such large-wavelength oscillations cannot be neglected and play a role of a constant noise. An instructive exercise is to compute the temperature that would be needed to produce $\bar n=354666$ if one took, instead of $2{\texttt{g}}$, the resonant frequency $\omega_0$: The result is $T=869770$K. The KMS condition then implies that the systems jumps up and down between the two states with practically identical rates, $\gamma_\downarrow\approx \gamma_\uparrow$.

The generalized model takes this transition into account but otherwise is the same as the one of Scala {\it et al\/}. The new master equation possesses two additional terms responsible for the long-wave open-cavity noise,
\begin{eqnarray}
\dot{\rho}
&=&%1
-i
[\Omega,\rho]
\nonumber
\\
&\pp=&
+
\gamma_1
\Big(
\frac 1 2
|\Omega_0\rangle \langle \Omega_+| \rho |\Omega_+\rangle \langle \Omega_0|
-
\frac 1 4
[|\Omega_+\rangle \langle \Omega_+|,\rho]_+
\Big)
\nonumber
\\
&\pp=&%2
+
\gamma_a
\Big(
\frac 1 2
|\Omega_+\rangle \langle \Omega_0| \rho |\Omega_0\rangle \langle \Omega_+|
-
\frac 1 4
[|\Omega_0\rangle \langle \Omega_0|,\rho]_+
\Big)
\nonumber
\\
&\pp=&%3
+
\gamma_2
\Big(
\frac 1 2
|\Omega_0\rangle \langle \Omega_-| \rho |\Omega_-\rangle \langle \Omega_0|
-
\frac 1 4
[|\Omega_-\rangle \langle \Omega_-|, \rho]_+
\Big)
\nonumber
\\
&\pp=&%4
+
\gamma_b
\Big(
\frac 1 2
|\Omega_-\rangle \langle \Omega_0| \rho |\Omega_0\rangle \langle \Omega_-|
-
\frac 1 4
[|\Omega_0\rangle \langle \Omega_0|,\rho]_+
\Big)
\nonumber
\\
&\pp=&%5
+
\gamma_3
\Big(
\frac 1 2
|\Omega_-\rangle \langle \Omega_+| \rho |\Omega_+\rangle \langle \Omega_-|
-
\frac 1 4
[|\Omega_+\rangle \langle \Omega_+|,\rho]_+
\Big)
\nonumber
\\
&\pp=&%6
+
\gamma_c
\Big(
\frac 1 2
|\Omega_+\rangle \langle \Omega_-| \rho |\Omega_-\rangle \langle \Omega_+|
-
\frac 1 4
[|\Omega_-\rangle \langle \Omega_-|,\rho]_+
\Big),
\label{eq:89}
\end{eqnarray}
involving an additional Davies operator $|\Omega_-\rangle\langle\Omega_+|$ and its Hermitian conjugate \cite{Alicki}. The decay parameters are explained in Fig.~\ref{fig:drabinka}.
Putting $\gamma_3=\gamma_c=0$ we reconstruct the model of Scala {\it et al.\/}
To solve (\ref{eq:89}) we employ the damping basis method \cite{Englert}. We begin with the time-independent operator ${\cal L}(\texttt{g})\rho(t)$ defined by the right side of (\ref{eq:89}), and solve its eigenvalue problem. The off-diagonal eigenvectors
\begin{subequations}
\begin{eqnarray}
{\cal{L}}(\texttt{g}) (|\Omega_+ \rangle \langle \Omega_-| )
&=&%1
\Big(
-i ( \Omega_+ - \Omega_- )
-
\frac{ \gamma_1 + \gamma_2 + \gamma_3 + \gamma_c }{ 4 }
\Big)
|\Omega_+\rangle \langle \Omega_-|,\\
\label{eq:3200}
{\cal{L}}(\texttt{g}) ( |\Omega_+\rangle \langle \Omega_0| )
&=&%1
\Big(
-i ( \Omega_+ - \Omega_0 )
-
\frac{ \gamma_1 + \gamma_3 + \gamma_a + \gamma_b  }{ 4 }
\Big)
|\Omega_+\rangle  \langle \Omega_0|
\label{eq:3201},\\
{\cal{L}}(\texttt{g})( |\Omega_-\rangle \langle \Omega_0| )
&=&%1
\Big(
-i ( \Omega_- - \Omega_0 )
-
\frac{ \gamma_2 + \gamma_a + \gamma_b + \gamma_c }{ 4 }
\Big)
|\Omega_-\rangle \langle \Omega_0|,
\label{eq:3202}
\end{eqnarray}
\end{subequations}
can be supplemented by their Hermitian conjugates corresponding to complex-conjugated eigenvalues.

To find the next three operators we solve the eigenequation
\begin{eqnarray}
{\cal{L}}(\texttt{g})
\Big(
x |\Omega_+\rangle \langle \Omega_+|
+
y |\Omega_-\rangle \langle \Omega_-|
+
z |\Omega_0\rangle \langle \Omega_0|
\Big)
&=&
\Lambda
\Big(
x |\Omega_+\rangle \langle \Omega_+|
+
y |\Omega_-\rangle \langle \Omega_-|
+
z |\Omega_0\rangle \langle \Omega_0|
\Big),
\end{eqnarray}
which is equivalent to
\begin{subequations}
\begin{eqnarray}
\frac 1 2
\big(
y \gamma_c
-
x ( \gamma_1  +  \gamma_3  )
+
z \gamma_a
\big)
&=&%1
\Lambda x,
\\
\frac 1 2
\big(
x \gamma_3
-
y ( \gamma_2 + \gamma_c )
+
z \gamma_b
\big)
&=&%2
\Lambda y,
\\
\frac 1 2
\big(
x \gamma_1
+
y \gamma_2
-
z ( \gamma_a  + \gamma_b  )
\big)
&=&%3
\Lambda z.
\end{eqnarray}
\end{subequations}
The solutions are
\begin{subequations}
\begin{eqnarray}
x_1&=& \gamma_b \gamma_c +  \gamma_a (\gamma_2 + \gamma_c),\\
y_1&=&\gamma_3 \gamma_a + \gamma_b (\gamma_1  + \gamma_3 ),\\
z_1&=&\gamma_2 \gamma_3 + \gamma_1 (\gamma_2 + \gamma_c ), \\
\Lambda_1&=&0,\\
\label{eq:3203}
x_2
&=&%1
(\gamma_1 - \gamma_3)(\gamma_c - \gamma_a) - (\gamma_1 + \gamma_3 ) (\gamma_1 - \gamma_2 + \gamma_3 - \gamma_b + S),
\\
y_2
&=&%2
(\gamma_1 + \gamma_3) (\gamma_3 - \gamma_b) + \gamma_3 (\gamma_2 - \gamma_a + \gamma_c + S ) - \gamma_1 \gamma_b ,
\\
z_2
&=&%3
- 2 \gamma_2 \gamma_3 + \gamma_1 (\gamma_1 - \gamma_2 + \gamma_3 + \gamma_a + \gamma_b - \gamma_c + S) ,
\\
\Lambda_2
&=&%4
- \frac{1}{4} (  \gamma_1 + \gamma_2 + \gamma_3  + \gamma_a  + \gamma_b  + \gamma_c + S),
\label{eq:3204}\\
x_3
&=&%1
(\gamma_1 - \gamma_3)(\gamma_c - \gamma_a) - (\gamma_1 + \gamma_3)(\gamma_1 - \gamma_2 + \gamma_3 - \gamma_b - S),
\\
y_3
&=&%2
(\gamma_1 + \gamma_3) (\gamma_3 - \gamma_b) + \gamma_3 (\gamma_2 - \gamma_a + \gamma_c - S) - \gamma_1 \gamma_b,
\\
z_3
&=&%3
- 2 \gamma_2 \gamma_3 + \gamma_1 (\gamma_1 - \gamma_2 + \gamma_3 + \gamma_a + \gamma_b - \gamma_c - S)
\\
\Lambda_3
&=&%4
- \frac{1}{4} (  \gamma_1  + \gamma_2  + \gamma_3  + \gamma_a + \gamma_b  + \gamma_c - S),
\label{eq:3205}
\end{eqnarray}
\end{subequations}
where we have defined
\begin{eqnarray}
S
&=&%1
\sqrt{(\gamma_1-\gamma_2+\gamma_3-\gamma_a-\gamma_b+\gamma_c)^2+4 ( \gamma_1 - \gamma_2 )
   (\gamma_a-\gamma_c)}
\label{eq:3206}
\end{eqnarray}
For the purpose of further reference  we list below the set of nine eigenoperators of ${\cal L}(\texttt{g})$, defined by the above equations
\begin{subequations}
\begin{eqnarray}
\rho_1
&=&%1
x_1 |\Omega_+\rangle \langle \Omega_+| + y_1 |\Omega_-\rangle \langle \Omega_-| + z_1 |\Omega_0\rangle \langle \Omega_0|,
\label{eq:3220}
\\
\rho_2
&=&%2
x_2 |\Omega_+\rangle \langle \Omega_+| + y_2 |\Omega_-\rangle \langle \Omega_-| + z_2 |\Omega_0\rangle \langle \Omega_0|,
\\
\rho_3
&=&%3
x_3 |\Omega_+\rangle \langle \Omega_+| + y_3 |\Omega_-\rangle \langle \Omega_-| + z_3 |\Omega_0\rangle \langle \Omega_0|,
\\
\rho_4
&=&%4
|\Omega_+\rangle \langle \Omega_-|,
\\
\rho_5
&=&%5
|\Omega_+\rangle \langle \Omega_0|,
\\
\rho_6
&=&%6
|\Omega_-\rangle \langle \Omega_0|,
\\
\rho_7
&=&%7
|\Omega_-\rangle \langle \Omega_+|,
\\
\rho_8
&=&%8
|\Omega_0\rangle \langle \Omega_+|,
\\
\rho_9
&=&%9
|\Omega_0\rangle \langle \Omega_-|.
\label{eq:3230}
\end{eqnarray}
\label{eq:3231}
\end{subequations}
We assume, as before, that the system is in exact resonance. The assumption has one extremely important technical implication: The eigenvectors $\rho_1$, $\dots$, $\rho_9$, are independent of $\texttt{g}$. The coupling parameter $\texttt{g}$ enters only into the ``off-diagonal" eigenvalues $\Lambda_4, \dots, \Lambda_9$. This property will allow us to derive an exact formula for $\rho(t)$, even for the Gaussian mode profile.

\begin{center}
\begin{figure}
\includegraphics{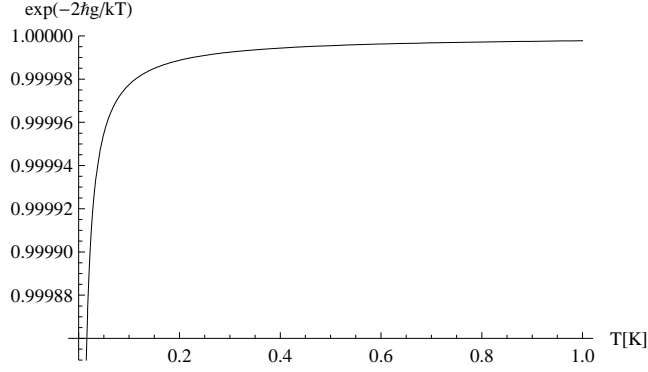}%
\caption{The dependence of the factor $e^{-\frac{2\hbar \texttt{g} }{ k T}}$ multiplying $\gamma_3$ in equation (\ref{eq:3280}) on temperature. For a wide range of temperatures we can simply set $\gamma_c=\gamma_3$; $\texttt{g}=47 \pi  10^3$Hz.}
\label{fig:gamma3gammac}
\end{figure}
\end{center}

Up to this point all calculations have been carried out without any assumption about the values of $\gamma_a$, $\gamma_b$ and $\gamma_c$. This strategy, however, will no longer pay as the resulting equations would be obscure and difficult to analyze.
Since at $T=0.8$K
\begin{subequations}
\begin{eqnarray}
\gamma_a
&=&%1
e^{-\frac{ \hbar (\omega_0 + {\tt{g}} ) }{ k T } }
\gamma_1
\approx
0.0466327
\gamma_1,
\\
\gamma_b
&=&%2
e^{-\frac{ \hbar (\omega_0 - {\tt{g}} ) }{ k T } }
\gamma_2
\approx
0.0466328
\gamma_2,
\\
\gamma_c
&=&%3
e^{-\frac{ 2 \hbar {\tt{g}}  }{ k T } }
\gamma_3
\approx
0.999997
\gamma_3,
\label{eq:3280}
\end{eqnarray}
\end{subequations}
it is justified to put $\gamma_a=\epsilon \gamma_1$, $\gamma_b=\epsilon \gamma_2$ with $\epsilon=0.0466$, and $\gamma_c=\gamma_3$. Fig.~\ref{fig:gamma3gammac} shows that $\gamma_c\approx\gamma_3$ is valid for a wide range of temperatures. For temperatures of the order of $10^{-3}$K the departures from exact equality amount to $0.003\gamma_3$.
Rewriting the initial condition $\rho(0)=|e,0\rangle \langle e,0|$ by means of the eigenvectors we use
\begin{eqnarray}
|\Omega_+\rangle \langle \Omega_+|
&=&%1
\frac{1}{  2 \epsilon + 1 }
\frac{1}{\gamma_2 \gamma_3 + \gamma_1 (\gamma_2 + \gamma_3) }
\rho_1
-
\frac{1}{2 \epsilon + 1}
\frac{ \gamma_1 + \gamma_2 + 2 \gamma_3 + \epsilon(\gamma_1 + \gamma_2) -  S   }{4 S (\gamma_2 \gamma_3 + \gamma_1 (\gamma_2 + \gamma_3))}
\rho_2
\nonumber
\\
&&%2
+
\frac{ 1 }{2\epsilon + 1 }
\frac{ \gamma_1 + \gamma_2 + 2 \gamma_3 + \epsilon (\gamma_1 + \gamma_2 ) +S   }{ 4 S(\gamma_2 \gamma_3 + \gamma_1 ( \gamma_2 + \gamma_3))}
\rho_3,
\ee
\be
|\Omega_-\rangle \langle \Omega_-|
&=&%1
\frac{ 1}{ 2\epsilon + 1}
\frac{ 1 }{ \gamma_2 \gamma_3 + \gamma_1 (\gamma_2 + \gamma_3 ) }
\rho_1
\nonumber
\\
&&%2
+
\frac{\gamma_1^2 (\epsilon +1)^2-S (\epsilon  \gamma_1+\gamma_1+\gamma_3)-\gamma_3 (3 \epsilon  \gamma_2+\gamma_2-2 \gamma_3)+\gamma_1 \left(-\epsilon  \gamma_3+\gamma_3+\gamma_2 \left(\epsilon ^2-2 \epsilon -1\right)\right)}{4 S (\gamma_2 \gamma_3 +\gamma_1 (\gamma_2+\gamma_3)) (2 \epsilon +1) (\gamma_1 \epsilon -\gamma_3)}
\rho_2
\nonumber
\\
&&%2
+
\frac
{
-\gamma_1^2 (\epsilon +1)^2
-
S (\gamma_1 + \gamma_3 + \epsilon  \gamma_1  )
+
\gamma_3 (\gamma_2 -2 \gamma_3 + 3 \epsilon  \gamma_2)
+
\gamma_1 \left(\epsilon  \gamma_3-\gamma_3+\gamma_2 \left(1 + 2 \epsilon - \epsilon ^2  \right)\right)}{4 S (\gamma_2
\gamma_3+\gamma_1 (\gamma_2+\gamma_3)) (2 \epsilon +1) (\gamma_1 \epsilon -\gamma_3)}
\rho_3.
\end{eqnarray}
Accordingly,
\begin{eqnarray}
\rho(0)
&=&%1
\frac{ 1  }{ 2\epsilon + 1 }
\frac{ 1 }{ \gamma_2 \gamma_3 + \gamma_1 (\gamma_2 + \gamma_3 ) } \rho_1
\nonumber
\\
&&%2
+
\frac 1 2
\frac{ 1 }{ 2 \epsilon + 1 }
\frac{ \gamma_1^2 (1+\epsilon ) - \gamma_1 \gamma_2 ( 1 + 3 \epsilon    ) + 2 \gamma_1 \gamma_3 ( 1 - \epsilon  )
+2 \gamma_3 (2 \gamma_3 - \gamma_2 \epsilon ) - S (\gamma_1+2 \gamma_3) }
{4 S (\gamma_2 \gamma_3 + \gamma_1 (\gamma_2+\gamma_3))
(\epsilon \gamma_1  -\gamma_3)}
\rho_2
\nonumber
\\
&&%3
-
\frac 1 2
\frac{ 1 }{ 2 \epsilon +1 }
\frac{ \gamma_1^2 ( 1 + \epsilon ) - \gamma_1 \gamma_2 ( 1+ 3 \epsilon   ) + 2 \gamma_1 \gamma_3 (1- \epsilon)
+ 2 \gamma_3 (2 \gamma_3 - \gamma_2 \epsilon ) + S ( \gamma_1 + 2 \gamma_3 ) }
{4 S (\gamma_2 \gamma_3+\gamma_1 (\gamma_2+\gamma_3)) ( \epsilon \gamma_1  -\gamma_3)}
\rho_3
\nonumber
\\
&&%4
-
\frac 1 2 \rho_4
-
\frac 1 2 \rho_7,
\label{eq:3250}\\
&=&%1
\sum_{i \in {\cal{I}} }
A_i \rho_i,
\qquad {\cal{I}}=\{1,2,3,4,7\},
\end{eqnarray}
the latter equality defining the coefficients $A_i$. Now
\begin{eqnarray}
\rho(t)
&=&
\sum_{j=1}^3
A_j e^{\Lambda_j t} \rho_j
+
A_4 e^{\Lambda_4(\texttt{g}) t} \rho_4
+
A_7 e^{\Lambda_7(\texttt{g}) t} \rho_7.
\label{eq:3041}
\end{eqnarray}
For $t\to\infty$ the system arrives at a finite-dimensional thermal equilibrium
\begin{eqnarray}
\lim_{t\to \infty}\rho(t)
&=&%1
\frac{ \epsilon }{ 2 \epsilon + 1} |\Omega_+\rangle \langle \Omega_+|
+
\frac{ \epsilon }{ 2 \epsilon + 1}
|\Omega_-\rangle \langle \Omega_-|
+
\frac{1}{ 2 \epsilon + 1}
|\Omega_0\rangle \langle \Omega_0|.
\label{eq:3350}
\end{eqnarray}
The coefficients multiplying $|\Omega_\pm\rangle \langle \Omega_\pm|$ are identical as a consequence of our simplifying assumption
$e^{-\frac{ \hbar (\omega_0 + {\tt{g}} ) }{ k T } }\approx e^{-\frac{ \hbar (\omega_0 - {\tt{g}} ) }{ k T } }\approx \epsilon$.
The fact that we defined Davies operators in the dressed-state basis results in $\rho(\infty)$ that commutes with the system Hamiltonian, as it should. Although we do not have an exact formula for $T>0$ in the phenomenological model, it seems that the asymptotic state will there commute with the free-field Hamiltonian, and thus not with $H$.

The asymptotic equilibrium is no longer a Gibbs state, a fact that follows from our truncated-space model. Finite dimensional thermal equilibria are known to occur in $q$-deformed statistics \cite{Tsallis}, so probably what we obtain is a state belonging to a $q$-exponential family \cite{Naudts1,Naudts2}, but the problem requires more detailed studies so we leave it open.

The explicit form of the probability we are interested in is finally
\begin{eqnarray}
p_g(t)
&=&%1
\frac { 1 + \epsilon }{ 1 + 2 \epsilon }
\nonumber
\\
&&%2
+
\frac{ 2 \gamma_3 - \epsilon (\gamma_1 + \gamma_2 ) -S }{4 S (2 \epsilon + 1)}
e^{ -  \frac{ \gamma_1 + \gamma_2 + 2 \gamma_3 + \epsilon (\gamma_1 + \gamma_2 )  +  S  }{ 4 }  t}
-
\frac{ 2 \gamma_3 - \epsilon (\gamma_1 + \gamma_2 ) + S    }{ 4 S (2 \epsilon + 1 ) }
e^{ - \frac { \gamma_1 + \gamma_2 + 2 \gamma_3 + \epsilon (\gamma_1 + \gamma_2 ) - S  }{  4  }  t }
\nonumber
\\
&&%3
-
\frac 1 2
e^{ - \frac { \gamma_1 + \gamma_2 + 2 \gamma_3  }{ 4 } t}
\cos 2 \texttt{g} t,
\label{eq:3290}
\end{eqnarray}
with the asymptotic value
\begin{eqnarray}
\lim_{t\to\infty} p_g(t)
&=&%1
\frac{1 + \epsilon }{ 1 + 2 \epsilon }
\approx
0.957.
\end{eqnarray}
A surprise is that if we put $T=0$, $\gamma_1=\gamma_2=0$, we find, for the first time in consequence of master-equation theoretical analysis, the $T=0$ fitting formula of Brune {\it et al.\/}.
\begin{eqnarray}
p_{g}(t)
=
\frac 1 2
-
\frac 1 2
e^{- \frac{ \gamma_3 }{ 4 }  t }
\cos 2 {\texttt{g}} t,\label{xxxx}
\end{eqnarray}
but with the damping parameter whose interpretation is completely different from what was expected.
This simple test suggests that the open-cavity model may be indeed the correct way towards explanation of the discussed experiment.
Let us remark that formulas analogous to (\ref{xxxx}) do occur in (semiclassical) Bloch-equation approaches to Rabi oscillations (cf. \cite{Walls}, especially Eq.~(11.4) and Fig.~11.2).

What yet has to be understood is the meaning of the damping parameters --- the next Section will be devoted to this problem.

Before we do that, let us briefly discuss the modification we get for the Gaussian profile. Repeating the argument we gave for the Scala {\it et al.\/} model, we find
\begin{center}
\begin{figure}
\includegraphics{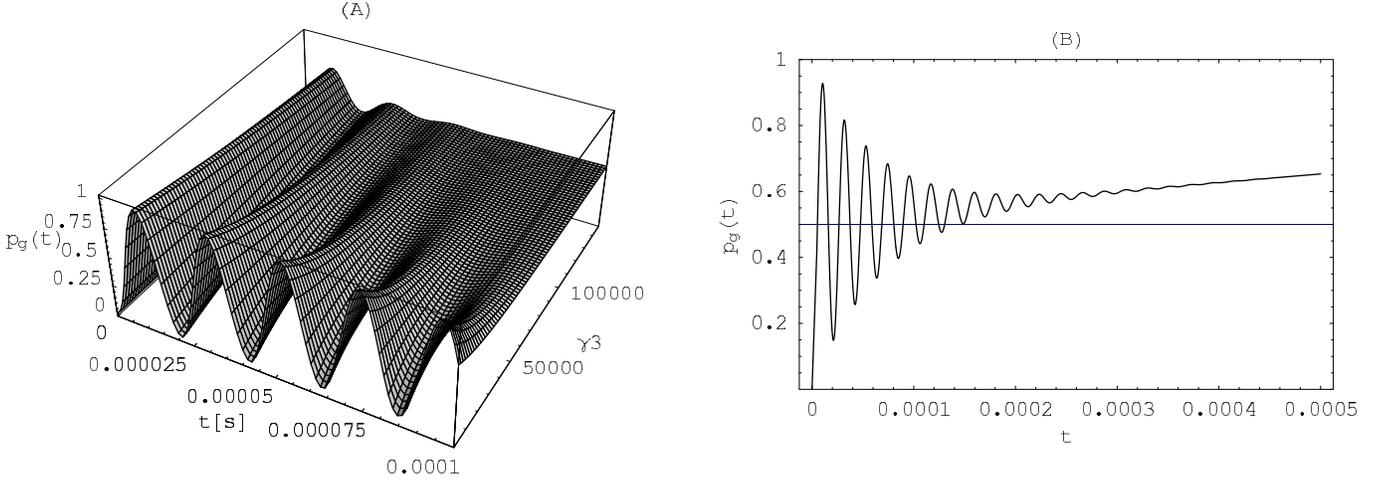}%
\caption{(A) Dependence of $p_g(t)$ on time and $\gamma_3$. The remaining parameters are $\texttt{g}=47 \pi 10^3$Hz and $\gamma_1=\gamma_2=0.01 \texttt{g}$. (B) $p_g(t)$ for a longer time (500 $\mu$s) for $\gamma_3=0.2 \texttt{g}$ and other parameters as in the left picture. $\gamma_3$ controls the rate at which oscillations decay towards $1/2$, whereas $\gamma_1$ and $\gamma_2$ are responsible for the rate at which $p_g(t)$ tends towards $1$, reaching it exactly only for $\epsilon=0$.}
\label{fig:zaleznosc-Pe-gamma3}
\end{figure}
\end{center}
\begin{eqnarray}
\rho(t)
&=&%1
A_1
e^{\Lambda_1 t } \    \rho_1
+
A_2
e^{\Lambda_2 t }\     \rho_2
+
A_3
e^{\Lambda_3 t }\     \rho_3
\nonumber
\\
&&%2
+
A_4
e^{  -\frac{ \gamma_1 + \gamma_2 + 2 \gamma_3 }{ 4 } t } \
e^{ -2 i {\texttt{g}} \sqrt{\pi } \frac w d  t } \
\rho_4
+
A_7
e^{  -\frac{ \gamma_1 + \gamma_2 + 2 \gamma_3 }{ 4 } t } \
e^{ 2 i \texttt{g} \sqrt{ \pi } \frac w d   t} \
\rho_7
\label{eq:3080}
\end{eqnarray}
and
\begin{eqnarray}
p_g(t)
&=&%1
\frac { 1 + \epsilon }{ 1 + 2 \epsilon }
\nonumber
\\
&&%2
+
\frac{ 2 \gamma_3 - \epsilon (\gamma_1 + \gamma_2 ) -S }{4 S (2 \epsilon + 1)}
e^{ -  \frac{ \gamma_1 + \gamma_2 + 2 \gamma_3 + \epsilon (\gamma_1 + \gamma_2 )  +  S  }{ 4 }  t}
-
\frac{ 2 \gamma_3 - \epsilon (\gamma_1 + \gamma_2 ) + S    }{ 4 S (2 \epsilon + 1 ) }
e^{ - \frac { \gamma_1 + \gamma_2 + 2 \gamma_3 + \epsilon (\gamma_1 + \gamma_2 ) - S  }{  4  }  t }
\nonumber
\\
&&%3
-
\frac 1 2
e^{ - \frac { \gamma_1 + \gamma_2 + 2 \gamma_3  }{ 4 } t}
\cos 2 \texttt{g}\sqrt{ \pi } \frac w d t.
\label{eq:3290'}
\end{eqnarray}%%
The result again agrees with the Brune {\it et al.\/} recipe, with all the restrictions we have made on the meaning of real and effective times and parameters. In practice, one also has to include the fact that real measurements involve some uncertainty of $t$ in $\rho(t)$, for example due to collisions of atoms with a background gas and the resulting velocity error.
An effective density matrix
\be
\rho_{\Delta t}(t)
&=&
\int_0^\infty dt'\, p_{\Delta t}(t,t')\rho(t'),
\ee
where \cite{Bonifacio}
\be
p_{\Delta t}(t,t')
&=&
\frac{e^{-t'/\Delta t}}{\Delta t}\frac{(t'/\Delta t)^{t/\Delta t-1}}{\Gamma(t/\Delta t)},
\ee
satisfies a collisional nondissipative master equation \cite{Bonifacio,Walls,Puri} with a double-commutator term $-\frac{\Delta t}{2}[\Omega,[\Omega,\rho_{\Delta t}(t)]]$.
The corresponding ground-state probability
\be
p_{g,\Delta t}(t)
&=&
\int_0^\infty dt'\, p_{\Delta t}(t,t')p_g(t')\label{p Delta t}
\ee
reconstructs the idealized case by $\lim_{\Delta t\to 0}p_{g,\Delta t}(t)=p_{g}(t)$.
Inserting (\ref{eq:3290'}) into (\ref{p Delta t}) we find
\be
p_{g,\Delta t}(t)
&=&
\frac { 1 + \epsilon }{ 1 + 2 \epsilon }
\nonumber
\\
&&%2
+
\frac{ 2 \gamma_3 - \epsilon (\gamma_1 + \gamma_2 ) -S }{4 S (2 \epsilon + 1)}
\Big(1+\frac{ \gamma_1 + \gamma_2 + 2 \gamma_3 + \epsilon (\gamma_1 + \gamma_2 )  +  S  }{ 4 }\Delta t\Big)^{-\frac{t}{\Delta t}}
\nonumber
\\
&&%2
-
\frac{ 2 \gamma_3 - \epsilon (\gamma_1 + \gamma_2 ) + S    }{ 4 S (2 \epsilon + 1 ) }
\Big(1+\frac { \gamma_1 + \gamma_2 + 2 \gamma_3 + \epsilon (\gamma_1 + \gamma_2 ) - S  }{  4  }\Delta t\Big)^{-\frac{t}{\Delta t}}
\nonumber
\\
&&%3
-
\frac 1 2
\Big[ \Big(1+ \frac { \gamma_1 + \gamma_2 + 2 \gamma_3  }{ 4 } \Delta t\Big)^2+\frac{4\pi\texttt{g}^2 w^2}{d^2}\Delta t^2\Big]^{-\frac{t}{2\Delta t}}
\cos \Big[\frac{t}{\Delta t}\arctan\Big( 2\texttt{g}\sqrt{ \pi } \frac w d \frac{\Delta t}{1+ \frac { \gamma_1 + \gamma_2 + 2 \gamma_3  }{ 4 } \Delta t}\Big)\Big].\label{effective pg}
\ee

\section{Cavity $Q$ factor versus damping parameters}

For a constant $\texttt{g}$ the average energy confined in the cavity would be  $\overline{E(t)}=\hbar\tr\Omega\rho(t)=\hbar\overline{\Omega(t)}$, where
\begin{eqnarray}
\overline{\Omega(t)}
&=&%1
\frac{ \omega_0 }{  2  }
\frac{ 2 \epsilon - 1 }{ 2 \epsilon + 1}
\nonumber
\\
&&%3
+
\frac{ \gamma_1 \epsilon (\omega_0 + 2 \texttt{g} ) + \gamma_2 \epsilon (\omega_0 - 2 \texttt{g}) + \texttt{g} (\gamma_1 - \gamma_2)  - 2  \omega_0  \gamma_3  + S  \omega_0    }{ 2 S ( 2\epsilon + 1 )  }
e^{ - \frac { \gamma_1 + \gamma_2 + 2 \gamma_3 + \epsilon (\gamma_1 + \gamma_2 ) + S }{ 4 }t }
\nonumber
\\
&&%4
-
\frac{ \gamma_1 \epsilon (\omega_0 + 2 \texttt{g} ) + \gamma_2 \epsilon (\omega_0 - 2 \texttt{g}) + \texttt{g} (\gamma_1 - \gamma_2 )  -2 \omega_0 \gamma_3  - S \omega_0   }{ 2 S (2\epsilon + 1)}
e^{ - \frac { \gamma_1 + \gamma_2 + 2 \gamma_3 + \epsilon (\gamma_1 + \gamma_2 ) - S }{ 4 }t }.
\label{eq:3310}
\end{eqnarray}
Asymptotically,
\begin{eqnarray}
\lim_{t\to\infty}
\Big(
\overline{\Omega(t)}
+
\frac{ \omega_0 }{ 2 }
\Big)
=
\omega_0
\Big(
1 - \frac{ 1 }{ 2 \epsilon + 1 }
\Big).
\end{eqnarray}
It should be kept in mind that  $\epsilon=0$ does not recover the $T=0$ case, since (\ref{eq:3310}) has been derived under the simplifying assumption that $\gamma_3=\gamma_c$, so that despite transitions $|\Omega_0\rangle \to |\Omega_\pm\rangle$ are not allowed, the atom-field system can still circulate between $|\Omega_+\rangle$ and $|\Omega_-\rangle$ with the rate determined by $\gamma_3$.

An interesting case arises if we assume that $\gamma_1 = \gamma_2 = \gamma$, when $\overline{\Omega(t)}$ becomes independent of $\gamma_3$, and reduces to
\begin{eqnarray}
\overline{\Omega(t)}
+
\frac{ \omega_0 }{ 2 }
&=&%1
\frac{ 2 \epsilon +  e^{- \frac{\gamma (2 \epsilon + 1)}{2} t}  }{ 2 \epsilon + 1 }
\omega_0.
\label{eq:3330}
\end{eqnarray}
This result can be interpreted as follows. Since $\gamma_1=\gamma_2$, both $|\Omega_+\rangle$ and $|\Omega_-\rangle$ are indistinguishable from the point of view of the dissipative processes. In other words, transitions between $|\Omega_-\rangle$ and $|\Omega_+\rangle$ do not influence the leaks of energy from the system. Moreover, at $T=0$ the decay simplifies even further
\begin{eqnarray}
\overline{\Omega(t)}+
\frac{ \omega_0 }{ 2 }
&=&%1
e^{ - \frac \gamma 2 t}
\omega_0
\label{eq:3320}
\end{eqnarray}
Coming back to (\ref{eq:3310}), with no assumptions about $\gamma_1$, $\gamma_2$, $\gamma_3$ and $\epsilon$, we can compare it to a  simple exponential energy decay,
\begin{eqnarray}
\overline{\Omega(t)}  -   \overline{\Omega(\infty)}
&\simeq  &%1
\Big(\overline{\Omega(0)}  -   \overline{\Omega(\infty)}\Big)
e^{ - \frac{ \omega_0 }{ Q }     t  }=\frac{ \omega_0 }{ 2 \epsilon + 1 }
e^{ - \frac{ \omega_0  }{ Q }  t  },\label{fitt}
\end{eqnarray}
where $Q$ is the cavity quality factor.

To find the optimal set of parameters $\gamma_1$, $\gamma_2$ and $\gamma_3$ we use the iterative Levenberg-Marquardt algorithm for nonlinear optimization, setting $\epsilon=0.0466$ ($T=0.8$K) and $Q=7\cdot 10^7$ as reported in \cite{Brune}. Straightforward numerical analysis shows that the best fit to the right side of (\ref{fitt}) is obtained for $\gamma_1=\gamma_2=1772$Hz  and arbitrary $\gamma_3$, which is shown in Fig.~\ref{fig:energy_decay}A. But when inserting these parameters into $p_g(t)$ we obtain poor agreement with the data (Fig.~\ref{fig:energy_decay}B; the best agreement was found for $\gamma_3=0.057\texttt{g}$). If, on the other hand, we try to fit the Rabi oscillation data (Fig.~\ref{fig:energy_decay}D), the optimal parameters  $\gamma_1=\gamma_2=17.73$Hz, $\gamma_3=0.07\texttt{g}$ imply $Q=3.31\cdot 10^{10}$ (Fig.~\ref{fig:energy_decay}C). The plots in Fig.~\ref{fig:energy_decay}A and Fig.~\ref{fig:energy_decay}C will not change if we replace in (\ref{eq:3310})
$\texttt{g}=47 \pi 10^3$Hz by $\texttt{g}=0$. The fits (\ref{fitt}) are thus insensitive to the exact form of $\texttt{g}(t)$ within the range of parameters typical of the experiment \cite{Brune}. Moreover, there is no visible change in the energy decay plots if we replace $\overline{\Omega(t)}$ by
\be
\overline{\Omega_{\Delta t}(t)}
&=&
\int_0^\infty dt'\, p_{\Delta t}(t,t')\overline{\Omega(t')},\label{Omega Delta t}
\ee
with $\Delta t$ even as large as $5\mu$s, which is greater than the time sampling step employed in the experiment. In consequence, the discrepancy between $Q$ we find and the parameter reported in the experiment cannot be explained by finite $\Delta t$.

\begin{center}
\begin{figure}
\includegraphics{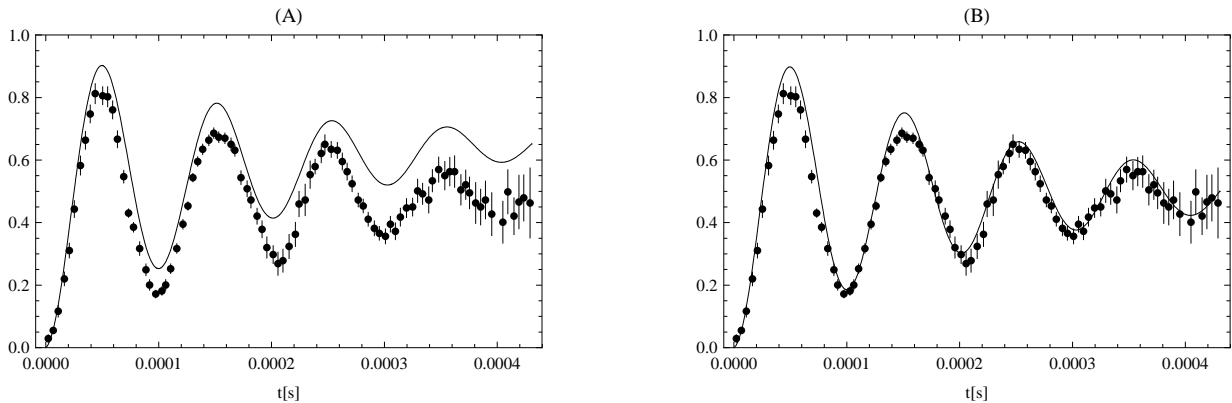}%
\caption{(A) The probability $p_{g,\Delta t}(t)$  given by (\ref{effective pg}) with  $\gamma_3=0$  and $\gamma_1=\gamma_2=1772$Hz, confronted with the Rabi oscillation data of \cite{Brune}. Parameters $\gamma_1$ and $\gamma_2$ are found in the procedure of fitting with exponential energy decay corresponding to $Q=7\cdot 10^7$, as reported in \cite{Brune}. (B) An analogous plots but with $Q=3.31\cdot 10^{10}$, $\gamma_1=\gamma_2=17.73$Hz, $\gamma_3=0$. $t$ is the true time and $\Delta t=2.37\mu$s, as estimated in \cite{Bonifacio}.}
\label{fig:Bonif}
\end{figure}
\end{center}

\begin{center}
\begin{figure}
\includegraphics{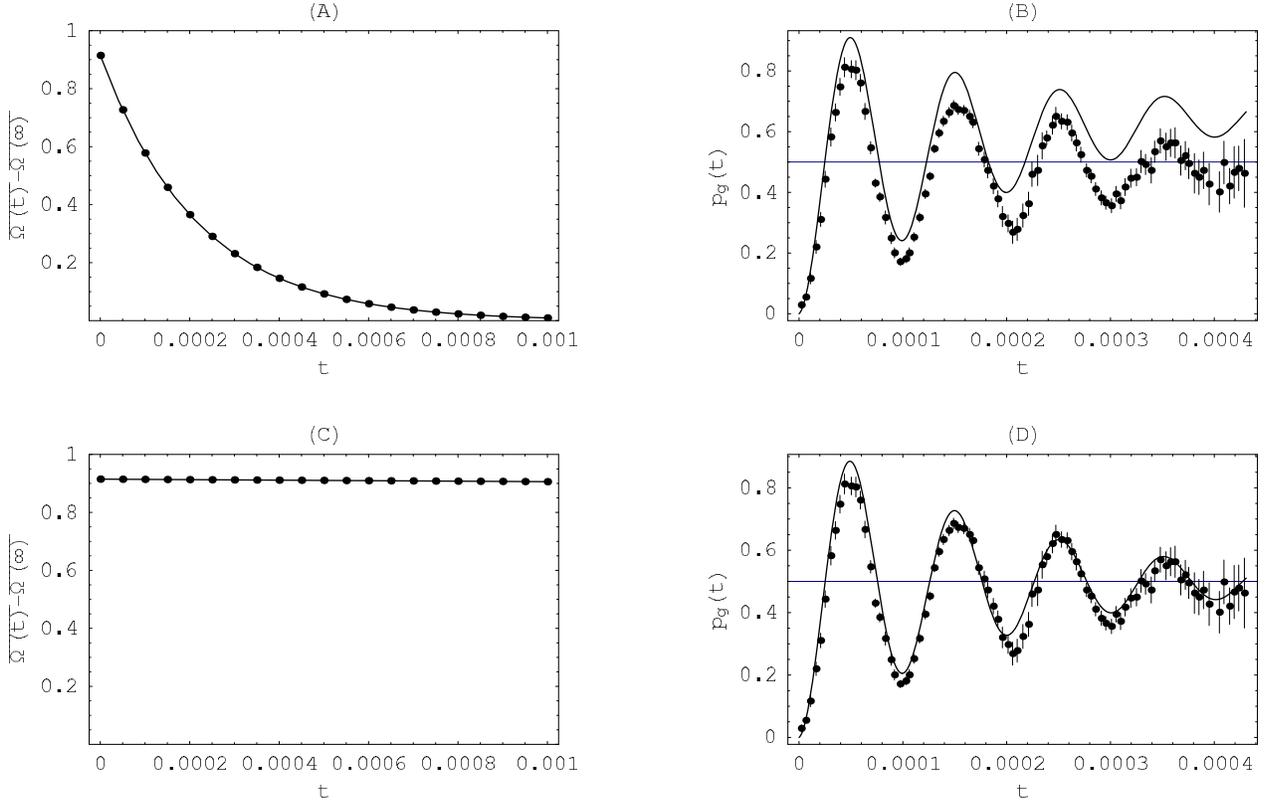}%
\caption{(A) The average $\overline{\Omega(t)}-\overline{\Omega(\infty)}$ of the atom-field system predicted by our model (solid) compared with the exponential decay (dots) modeled by $\big(\overline{\Omega(0)}  -   \overline{\Omega(\infty)}\big)
e^{ - \frac{ \omega_0 }{ Q }     t  }$, with $Q=7\cdot 10^7$ as reported in \cite{Brune}, and $\gamma_1=\gamma_2=1772$Hz found by optimization. (B) The probability $p_g(t)$  predicted according to (\ref{eq:3290'}) with $\gamma_1=\gamma_2=1772$Hz, $\gamma_3=0.07\texttt{g}$ found on the basis of the reported $Q$ factor, confronted with the Rabi oscillation data of \cite{Brune}. Parameters $\gamma_1$ and $\gamma_2$ are found in the procedure of fitting with a simple exponential energy decay, while $\gamma_3$ minimizes the sum of squared errors. (C) and (D) show analogous plots, but with $Q=3.31\cdot 10^{10}$, $\gamma_1=\gamma_2=17.73$Hz, $\gamma_3=0.07\texttt{g}$; $t$ is the true time. Various combinations of $\gamma_3$ and $\Delta t$ lead to practically identical plots for $p_{g,\Delta t}(t)$ (e.g. $\Delta t=0.5\mu$s, $\gamma_3=0.057\texttt{g}$).}
\label{fig:energy_decay}
\end{figure}
\end{center}

Let us now compare the data from \cite{Brune} with (\ref{effective pg}) under the assumption that $\gamma_3=0$, that is if the long-wave transition $|\Omega_+\rangle \leftrightarrow |\Omega_-\rangle$ is absent.
The estimates made in \cite{Bonifacio} led, for the experiment of Brune {\it et al.\/}, to the error $\Delta t\approx  0.5\mu$s of the {\it effective\/} time. For the true time the corresponding value is $\Delta t\approx  2.37\mu$s. Inserting the latter value into (\ref{effective pg}) and assuming $\gamma_3=0$ we obtain damped Rabi oscillations shown in Fig.~~\ref{fig:Bonif}. The agreement is again good.

The conclusion is that the effects described in \cite{Bonifacio} may be of the same order as those arising from opening the cavity. Since the cavity {\it was\/} in fact open, it seems that the realistic value of $\Delta t$ was probably smaller than the one discussed in \cite{Bonifacio}.
\section{Dynamics of entanglement between atomic and photon degrees of freedom}

The last issue we want to address is for how long the atoms and photons in the cavity can be regarded as an inseparable bipartite system \cite{Werner}. The problem is important for cavity QED implementations of quantum information processing, and is not entirely obvious since we know that there exist two decoherence time scales. Energy losses are characterized by $\gamma_1$ and $\gamma_2$ (damping of $p_g(t)$ towards 1), while $\gamma_3$ is responsible for the nondissipative damping of $p_g(t)$ towards $1/2$. Supplementing the three basis bare states, $|10\rangle=|e,0\rangle$, $|01\rangle=|g,1\rangle$, $|00\rangle=|g,0\rangle$, by the fourth state $|11\rangle=|e,1\rangle$ we obtain a bipartite $2\times 2$ composite system whose separability is completely characterized by the Peres-Horodecki partial transposition criterion \cite{Peres,HHH}.

So, let us take the solution $\rho(t)$, given by (\ref{eq:3080}) but expressed as a matrix of coefficients evaluated in the basis of bare states, and embed it in the space of $4\times 4$ matrices,
\be
\rho(t)
=
\left(
\begin{array}{cccc}
0 & 0 & 0 & 0\\
0 & \rho_{10,10} & \rho_{10,01} & 0\\
0 & \rho_{01,10} & \rho_{01,01} & 0\\
0 & 0 & 0 & \rho_{00,00}
\end{array}
\right).
\ee
The Peres-Horodecki criterion states that $\rho(t)$ is separable if and only if the partially transposed matrix
\be
r(t)
=
\left(
\begin{array}{cccc}
0 & 0 & 0 & \rho_{10,01}\\
0 & \rho_{10,10} & 0  & 0\\
0 & 0 & \rho_{01,01} & 0\\
\rho_{01,10} & 0 & 0 & \rho_{00,00}
\end{array}
\right)
\ee
has a non-negative spectrum. It turns out that $r(t)$ corresponding to (\ref{eq:3080}) has the following eigenvalues
\be
\lambda_1
&=&
\langle g,1| \rho(t) |g,1\rangle\geq 0,
\\
\lambda_2
&=&
\langle e,0| \rho(t) |e,0\rangle\geq 0,
\\
\lambda_3
&=&
\frac 1 2 \left(  \langle g,0| \rho(t) |g,0\rangle + \sqrt{ \langle g,0| \rho(t) |g,0\rangle^2
+ 4 |\langle e,0| \rho(t) |g,1\rangle|^2}  \right)\geq 0,
\\
\lambda_4
&=&
\frac 1 2 \left(  \langle g,0| \rho(t) |g,0\rangle - \sqrt{ \langle g,0| \rho(t) |g,0\rangle^2
+ 4 |\langle e,0| \rho(t) |g,1\rangle|^2}  \right)\leq 0,
\ee
where
\be
\langle e,0| \rho(t) |g,1\rangle
&=&
\frac 1 4
e^{- \frac{ \gamma_2 }{ 2 } t}(\gamma_1 - \gamma_2 + 2 \gamma_3)
\frac{  e^{- \frac{ \gamma_1 -\gamma_2 + \gamma_3 }{ 2 } t}-1}{ \gamma_1 - \gamma_2 + \gamma_3}
+
\frac i 2
e^{- \frac{ \gamma_1 + \gamma_2 + \gamma_3 }{4} t}
\sin 2  \texttt{g} t.
\ee
The effective measure of entanglement in the system is thus
the value of
$
|\langle e,0| \rho(t) |g,1\rangle|=|\rho_{10,01}(t)|
$
which vanishes if and only if
$\gamma_1 - \gamma_2 + 2 \gamma_3=0$ and $\sin 2  \texttt{g} t=0$.

\begin{center}
\begin{figure}
\includegraphics[width=1\textwidth]{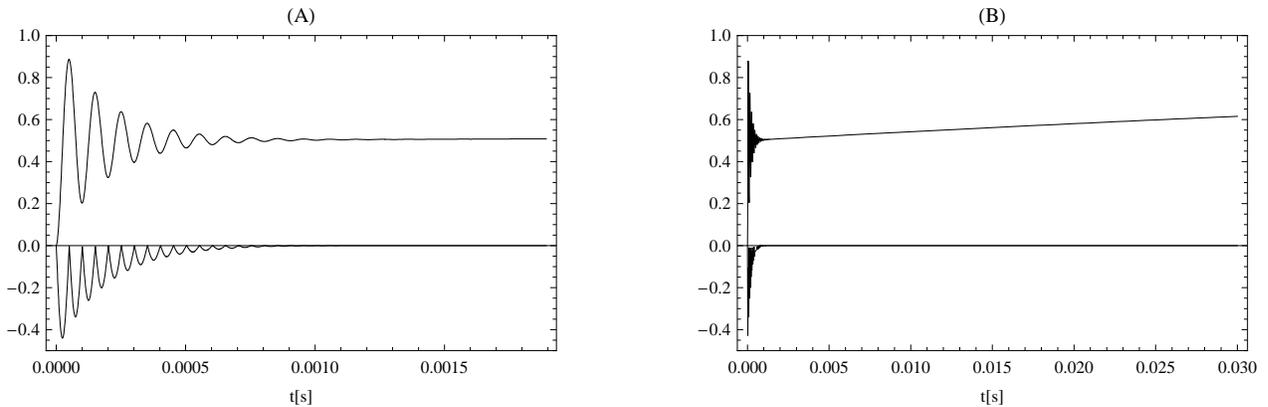}%
\caption{Dynamics of separability of $\rho(t)$. (A) The upper curve represents $p_g(t)$, the same as in Fig.~\ref{fig:energy_decay}D. The lower curve is the non-positive eigenvalue $\lambda_4$ of partially transposed $\rho(t)$. (B) The same curves as in (A) but for a longer time. Entanglement decays according to the value of $\gamma_3$ and is unrelated to the time scale of energy losses, controlled by $\gamma_1$ and $\gamma_2$.}
\label{fig:r}
\end{figure}
\end{center}

Fig.~\ref{fig:r} shows that energy dissipation $|\Omega_\pm\rangle\to|\Omega_0\rangle$ does not really influence separability properties of atom-photon density matrices. The crucial parameter is the one that controls fluctuations between $|\Omega_+\rangle$ and  $|\Omega_-\rangle$. The fact that it is the non-dissipative decoherence that plays the crucial role for quantum computational issues agrees with conclusions found by other means in \cite{Bonifacio2}. One can also think of it as consequence of continuous monitoring of the qubit by the noise arriving from outside of the cavity.

\section{Conclusions}

The formalism where Davies jump operators are defined in the dressed-state basis is mathematically simpler from the standard one where the jumps are defined by photonic creation and annihilation operators. Simultaneously, when extended to include the long-wave transitions typical of open cavities, the model predicts corrections to Rabi oscillations of the same order as those resulting from uncertainties in measurement of $t$. The so far ignored fluctuations $|\Omega_+\rangle\leftrightarrow |\Omega_-\rangle$ turn out to have fundamental implications for coherence properties of the atom--photon system in cavity QED. In particular, these are the transitions that crucially influence separability properties of states in the cavity and, hence, the attempts of cavity QED implementations of quantum gates have to minimize both  $\Delta t$ and $\gamma_3$. Improvements of cavity $Q$ factors as well as lowering the cavity temperature yet lower than $0.8$K may be less essential. The fact that the value of $Q$ we have determined on the basis of the data is some 500 times larger from the value reported in the experiment requires further studies. One of the possible interpretations is that the rate at which energy is lost in the cavity depends on $a^{\dag}a$, and thus for small light intensities the value of $Q$ may be much larger from the value one would have found for larger numbers of photons. This interpretation is supported by the observation \cite{Alicki} that the missing Davies operator can be regarded as a consequence of a $a^{\dag}a\otimes R$ system--reservoir interaction, where $R$ is some operator acting in the reservoir Hilbert space.

We are now in position to apply the above results to the problem of cavity QED tests of different field quantization paradigms, in particular to the issue of certain reducible representations of the algebra of quantum fields. There are reasons to believe that fields quantized in such representations are much less singular than the standard operator-valued distributions, so the question is directly related to various very fundamental issues that have not been subject to quantum optical tests as yet. At the level of vacuum Rabi oscillations the reducibility of representations should be manifested in terms of collapses and revivals in exact vacuum \cite{Czachor}, but observability of the effect --- it it really exists --- will crucially depend on controllability of the coherence losses plaguing cavity QED. These and related issues will be discussed in the second part of the present paper \cite{WCpart2}.

\section*{Appendix: Microscopic derivation of the open-cavity master equation}

The derivation presented in this Appendix was suggested to us by R. Alicki.
Let us start with the Hamiltonian
\begin{eqnarray}
H = \hbar\Omega_S + \hbar\Omega_R + \hbar\Omega_{I},
\label{eq:0}
\end{eqnarray}
where $\hbar\Omega_S$ and $\hbar\Omega_R$ are Hamiltonians of the system and the reservoir, respectively. The system-reservoir interaction is assumed in the form
\begin{eqnarray}
\Omega_I
&=&%1
A \otimes B
\ = \
\Big\{ \alpha \big( a+a^\dagger \big)  +  \beta a^\dagger a \Big\} \otimes \sum_k \texttt{g}_k (b_k + b^\dagger_k),
\label{eq:1}
\end{eqnarray}
($\alpha$ and $\beta$ are parameters). The term proportional to $\alpha$ is the standard one. The additional part, proportional to $\beta$, is the new term proposed by Alicki. Following the procedure described in \cite{KMS} we perform the decomposition of the operator $A$:
\be
A
&=&%1
\sum_\omega A(\omega),\\
A(\omega)
&=&%1
\sum_{\epsilon'-\epsilon=\omega}
\Pi(\epsilon) A \Pi(\epsilon').
\ee
$\Pi(\epsilon)$ is the projector on the eigensubspace corresponding to the eigenvalue $E=\hbar\epsilon$ of $H_S$. Then
\begin{eqnarray}
\big [\Omega_S,A(\omega) \big]
&=&%1
- \omega A(\omega),
\label{eq:71}
\\
\big[ \Omega_S,A^\dagger(\omega) \big]
&=&%2
+ \omega A^\dagger(\omega).
\label{eq:72}
\end{eqnarray}
and $A^\dagger(\omega)=A(-\omega)$.
Following the standard steps \cite{KMS} we arrive at
\begin{eqnarray}
\dot{\rho}(t)
&=&%1
-i [H_S,\rho(t)]
+
\sum_{\omega>0} \gamma(\omega)
\left(
A(\omega) \rho(t) A^\dagger(\omega)
-
\frac 1 2
\big[
A^\dagger(\omega) A(\omega), \rho(t)
\big]_+
\right)
\nonumber
\\
&&%2
+
\sum_{\omega>0} \gamma(-\omega)
\left(
A^\dagger(\omega) \rho(t) A(\omega)
-
\frac 1 2
\big[
A(\omega) A^\dagger(\omega), \rho(t)
\big]_+
\right),
\label{eq:50}
\end{eqnarray}
where we have ignored the energy shifts caused by the environment and assumed $\gamma(0)=0$ (which is true for the usual models of thermal reservoirs \cite{KMS}).

Now assume that $H_S$ is the Jaynes-Cummings Hamiltonian in exact resonance. The eigenstates of $H_S$ are the dressed states
\begin{eqnarray}
|\Omega_{N,\pm} \rangle
&=&%1
\frac{ 1 }{ \sqrt{2} }
\big(
|g,N\rangle \pm |e,N-1\rangle
\big).
\end{eqnarray}
The annihilation operators for the atom-field system are given explicitly by
\begin{eqnarray}
A(\Omega_{N',m'} - \Omega_{N,m} )
&=&%1
\Pi(\Omega_{N,m}) \Big\{  \alpha \big( a+ a^\dagger \big) +  \beta a^\dagger a \Big\} \Pi(\Omega_{N',m'})
\nonumber\\
&=&%11
\frac 1 2
\alpha
\left(
\sqrt{N+1}
+{mm'\,}
\sqrt{N}
\right)
\delta_{N',N+1}
|\Omega_{N,m} \rangle \langle \Omega_{N+1,m'} |
\nonumber
\\
&&%12
+
\frac 1 2
\alpha
\left(
\sqrt{N}
+{mm'\,}
\sqrt{N-1}
\right)
\delta_{N',N-1}
|\Omega_{N,m} \rangle \langle \Omega_{N-1,m'} |
\nonumber
\\
&&%13
+
\frac 1 2
\beta
\Big(
N
+{mm'\,}
(N-1)
\Big)
\delta_{N',N}
|\Omega_{N,m} \rangle \langle \Omega_{N,m'} |.
\label{eq:60}
\end{eqnarray}
To proceed further let us consider the two special cases, $N=0$ and $N=1$. For $N=0$,
\begin{eqnarray}
A(\Omega_{1,\pm }-\Omega_0)
&=&%1
\frac{ 1 }{ \sqrt{2} }
\alpha
|\Omega_{0} \rangle \langle \Omega_{1,\pm} |.
\label{eq:40}
\end{eqnarray}
For $N=1$ we get
\begin{eqnarray}
A(\Omega_{N',m'} -  \Omega_{1,m})
&=&%1
\frac 1 2
\alpha\
\delta_{N',2}
\left(
\sqrt{2}
+{mm'\,}
1
\right)
|\Omega_{1,m} \rangle \langle \Omega_{2,m'} |
\nonumber
\\
&&%2
+
\frac{ 1 }{ \sqrt{2} }
\alpha \
\delta_{N',0}
|\Omega_{1,m} \rangle \langle \Omega_{0} |
\nonumber
\\
&&%3
+
\frac 1 2
\beta
\delta_{N',1}
|\Omega_{1,m} \rangle \langle \Omega_{1,m'} |.
\end{eqnarray}
Keeping $N=1$ and substituting explicit values for $N'$, $m$ and $m'$ we get the following operators:
\be
A(\Omega_{2,+} - \Omega_{1,+})
&=&%1
\frac 1 2 \alpha ( \sqrt{2} + 1  )  |\Omega_{1,+}\rangle \langle \Omega_{2,+}|,
\\
A(\Omega_{2,+} - \Omega_{1,-})
&=&%2
\frac 1 2 \alpha ( \sqrt{2} - 1  )  |\Omega_{1,-}\rangle \langle \Omega_{2,+}|,
\\
A(\Omega_{2,-} - \Omega_{1,+})
&=&%3
\frac 1 2 \alpha ( \sqrt{2} - 1  )  |\Omega_{1,+}\rangle \langle \Omega_{2,-}|,
\\
A(\Omega_{2,-} - \Omega_{1,-})
&=&%4
\frac 1 2 \alpha ( \sqrt{2} + 1  )  |\Omega_{1,-}\rangle \langle \Omega_{2,-}|,
\\
A(\Omega_0 - \Omega_{1,\pm})
&=&%1
\frac{ 1 }{ \sqrt{2} } \alpha |\Omega_{1,\pm} \rangle \langle \Omega_0|,
\\
A(\Omega_{1,+} - \Omega_{1,+})
&=&%1
\frac 1 2 \beta |\Omega_{1,+} \rangle \langle \Omega_{1,+} |,
\\
A(\Omega_{1,+} - \Omega_{1,-})
&=&%2
\frac 1 2 \beta |\Omega_{1,-} \rangle \langle \Omega_{1,+} |,
\\
A(\Omega_{1,-} - \Omega_{1,+})
&=&%3
\frac 1 2 \beta |\Omega_{1,+} \rangle \langle \Omega_{1,-} |,
\\
A(\Omega_{1,-} - \Omega_{1,-})
&=&%4
\frac 1 2 \beta |\Omega_{1,-} \rangle \langle \Omega_{1,-} |.
\ee
In the simplest case $T=0$, and with the initial condition we have worked with above, we find
\begin{eqnarray}
\dot{\rho}(t)
&=&%1
-i [\Omega_S,\rho(t)]
\nonumber
\\
&&%2
+
\gamma(\Omega_{1,+}-\Omega_0)
\alpha^2
\left(
\frac 1 2
|\Omega_0\rangle \langle \Omega_+ | \rho (t)  |\Omega_+\rangle \langle \Omega_0|
-
\frac 1 4
\big[
|\Omega_+\rangle \langle \Omega_+|, \rho(t)
\big]_+
\right)
\nonumber
\\
&&%3
+
\gamma(\Omega_{1,-}-\Omega_0)
\alpha^2
\left(
\frac 1 2
|\Omega_0\rangle \langle \Omega_- | \rho (t)  |\Omega_-\rangle \langle \Omega_0|
-
\frac 1 4
\big[
|\Omega_-\rangle \langle \Omega_-|, \rho(t)
\big]_+
\right)
\nonumber
\\
&&%4
+
\gamma(\Omega_{1,+}-\Omega_{1,-})
\frac{ \beta^2 }{ 2 }
\left(
\frac 1 2
|\Omega_-\rangle \langle \Omega_+ | \rho (t)  |\Omega_+\rangle \langle \Omega_-|
-
\frac 1 4
\big[
|\Omega_+\rangle \langle \Omega_+|, \rho(t)
\big]_+
\right)
\nonumber.
\end{eqnarray}
And this is the equation we have postulated, if $\gamma_1=\gamma(\Omega_{1,+}-\Omega_0)\alpha^2$, $\gamma_2=\gamma(\Omega_{1,-}-\Omega_0)\alpha^2$,
and $\gamma_3=\gamma(\Omega_{1,+}-\Omega_{1,-})\beta^2/2$. The extension to $T>0$ is obvious.
\acknowledgments
The work was supported by VUB, Brussels, and Polish national scientific network LFPPI. We are indebted to Robert Alicki, Pawe{\l} Horodecki, Ryszard Horodecki, Matteo Scala and David Vitali for important remarks.

\end{document}